\documentclass[onecolumn, draftclsnofoot, 12pt]{IEEEtran}
\usepackage[T1]{fontenc}
\usepackage[latin9]{inputenc}
\usepackage[active]{srcltx}
\usepackage{color}
\usepackage{array}
\usepackage{verbatim}
\usepackage{float}
\usepackage{amsmath}
\usepackage{amsthm}
\usepackage{amssymb}
\usepackage{graphicx}

\makeatletter

\providecommand{\tabularnewline}{\\}
\floatstyle{ruled}
\newfloat{algorithm}{tbp}{loa}
\providecommand{\algorithmname}{Algorithm}
\floatname{algorithm}{\protect\algorithmname}

\theoremstyle{plain}
\newtheorem{thm}{\protect\theoremname}
\theoremstyle{plain}
\newtheorem{lem}[thm]{\protect\lemmaname}
\theoremstyle{remark}
\newtheorem{rem}[thm]{\protect\remarkname}

\usepackage{algorithmic}
\usepackage{cite}
\usepackage{diagbox}
\floatname{algorithm}{Algorithm}

\DeclareMathOperator{\tr}{tr}
\DeclareMathOperator{\minimize}{minimize}
\DeclareMathOperator{\diag}{diag}
\DeclareMathOperator{\maximize}{maximize}
\DeclareMathOperator{\st}{subject~to}
\newcommand{\herm}{^{{\dagger}}}
\newcommand{\trans}{^{\mathsf{T}}}

\usepackage{subfig}

\@ifundefined{showcaptionsetup}{}{%
 \PassOptionsToPackage{caption=false}{subfig}}
\usepackage{subfig}
\makeatother

\providecommand{\lemmaname}{Lemma}
\providecommand{\remarkname}{Remark}
\providecommand{\theoremname}{Theorem}

\begin{document}
\title{On the Secrecy Capacity of MIMO Wiretap Channels: Convex Reformulation
and Efficient Numerical Methods}
\author{Anshu Mukherjee,~\IEEEmembership{Student Member, IEEE}, Björn Ottersten,
~\IEEEmembership{Fellow, IEEE}, and Le-Nam Tran,~\IEEEmembership{Senior Member, IEEE}\thanks{Parts of this paper was presented at the IEEE Vehicular Technology
Conference (Spring), Helsinki, Finland, April 2021 \cite{Anshu:MIMOWTC:21VTC}.}\thanks{A. Mukherjee and L.-N. Tran are with School of Electrical and Electronic
Engineering, University College Dublin, Ireland. Email: anshu.mukherjee@ucdconnect.ie;nam.tran@ucd.ie.}\thanks{B. Ottersten is with Interdisciplinary Centre for Security, Reliability
and Trust, University of Luxembourg, Luxembourg. Email: bjorn.ottersten@uni.lu.}}
\maketitle
\begin{abstract}
This paper presents novel numerical approaches to finding the secrecy
capacity of the multiple-input multiple-output (MIMO) wiretap channel
subject to multiple linear transmit covariance constraints, including
sum power constraint, per antenna power constraints and interference
power constraint. An analytical solution to this problem is not known
and existing numerical solutions suffer from slow convergence rate
and/or high per-iteration complexity. Deriving computationally efficient
solutions to the secrecy capacity problem is challenging since the
secrecy rate is expressed as a difference of convex functions (DC)
of the transmit covariance matrix, for which its convexity is only
known for some special cases. In this paper we propose two low-complexity
methods to compute the secrecy capacity along with a convex reformulation
for degraded channels. In the first method we capitalize on the accelerated
DC algorithm which requires solving a sequence of convex subproblems,
for which we propose an efficient iterative algorithm where each iteration
admits a closed-form solution. In the second method, we rely on the
concave-convex equivalent reformulation of the secrecy capacity problem
which allows us to derive the so-called partial best response algorithm
to obtain an optimal solution. Notably, each iteration of the second
method can also be done in closed form. The simulation results demonstrate
a faster convergence rate of our methods compared to other known solutions.
We carry out extensive numerical experiments to evaluate the impact
of various parameters on the achieved secrecy capacity.

\textit{Index Terms} - MIMO, wiretap channel, secrecy capacity, sum
power constraint, per antenna power constraint, convex-concave.
\end{abstract}

\section{Introduction}

Security has always been a great concern to the public since the very
early days of wireless communications. This problem is increasingly
important nowadays since wireless connectivity becomes an integral
part of our modern life. Our dependency on wireless communications
and the associated risks are apparent during the current global pandemic.
Wireless communications offers great flexibility and convenience to
manage our lives but, at the same time, also creates more entries
for adversaries to attack due to its open broadcasting nature.

Primary methods for data security are traditionally based on cryptographic
techniques which are mainly implemented at the higher layers (e.g
the application layer) of the open systems interconnection (OSI) model
of a communication network. The computational complexity of the encryption
key management in cryptographic techniques is a major issue to apply
them to a large number of low-cost IoT devices or in dynamic and open
environments. Specifically, IoT devices are typically limited in terms
of storage and computing capability to handle such complicated encryption/decryption
algorithms. As a result, secure data transmission strategies based
on the physical properties of radio channels have been studied as
a promising alternative, which gives rise to physical layer security
(PLS). Compared to conventional cryptographic methods, PLS has distinguishing
advantages, including low-complexity in nature and possibly key-less
secure transmission. Thus, PLS is a powerful solution to address the
data security in future wireless networks, and has become a research
area of growing interest in the last decade.

The PLS started with the notion of perfect secrecy by Shannon in \cite{Shannon}.
In \cite{Wyner75}, Wyner introduced and studied the secrecy capacity
of the wiretap channel (WTC) which is a fundamental information-theoretic
model for PLS. In the WTC, a legitimate transmitter wishes to securely
transmit data to a legitimate receiver in the presence of an eavesdropper.
The secrecy capacity is simply thought as the maximum rate at which
the transmitter can reliably communicate with the receiver while ensuring
that the eavesdropper cannot decode the information. Since Wyner's
seminal paper, the WTC has been extended, covering various scenarios.
In particular, the secrecy capacity of the Gaussian WTC was studied
\cite{Gauss_wiretap}. The use of multiple antennas at transceivers
in contemporary wireless communications systems gives rise to the
so-called multiple-input multiple-output (MIMO) Gaussian WTC. The
secrecy capacity of the MIMO Gaussian WTC has received significant
interest since the late 2000s. In this regard, there have been many
results in the literature and we attempt to comprehensively (but by
no means exhaustively) summarize the significant ones below.

An analytical solution for the multiple-input single-output (MISO)
Gaussian WTC where both the eavesdropper and the legitimate receiver
have a single antenna was proposed in \cite{ZangLi}. When the channel
state information is perfectly known, the secrecy capacity of MIMO
WTC was characterized in \cite{Secrecy_cap_MISOME,MIMOME_WTC,Oggier2011SecCapEq}.
Particularly, explicit expressions for optimal signaling for MIMO
WTC are possible under some special cases \cite{Li2010MIMOwiretap,Sec_MIMO_SPC_3,Loyka2016}.
For example, in \cite{Sec_MIMO_SPC_3} Fakoorian \textit{et.al} presented
a full-rank solution for Gaussian MIMO WTC under an average power
constraint. A closed-form solution for optimal signaling for strictly
degraded Gaussian MIMO WTC with sufficiently large power was presented
in \cite{Loyka2016}. In \cite{MIMO_POTDC} the secrecy rate maximization
problem for the MIMO WTC with multiple eavesdroppers was considered,
in which an alternating matrix based algorithm, named polynomial time
difference of convex functions (POTDC), was introduced. Power minimization
and secrecy rate maximization for the MIMO WTC was studied in \cite{Cumanan2014}
using a difference of convex functions algorithm (DCA). More recently,
an efficient low-complex solution for the MIMO WTC was proposed in
\cite{ThangNguyen2020} using a convex-concave optimization framework.
We note that the above studies on MIMO wiretap channels focused on
the secrecy capacity subject to a sum power constraint (SPC).

In a MIMO communications system, each transmit antenna can be equipped
with a separate RF chain, and thus a per-antenna power constraint
(PAPC) is more practically relevant than constraining the sum power
\cite{MIMO_Cap_LTCC},\cite{Revisitng_MIMO_Cap}. In this regard,
the secrecy capacity with joint SPC and PAPC has been studied in \cite{Li2013,lyoka2015minmaxEQ}
for the Gaussian MIMO WTC, and more recently in \cite{Cao2020} for
the MISO Gaussian WTC. More specifically, an iterative algorithm that
combines the alternating optimization (AO) method and the subgradient
method to compute the secrecy capacity of the MIMO WTC for joint SPC
and PAPC was proposed in \cite{Li2013}. Based on an equivalent minimax
reformulation of the secrecy capacity problem, a barrier method was
presented in \cite{lyoka2015minmaxEQ}. Very recently, a closed-form
solution of optimal transmit strategies for Gaussian MISO wiretap
channels were derived in \cite{Cao2020}. In \cite{Dong2018c,Loyka2018},
the Gaussian MIMO WTC was extended to scenarios where the transmitter
also needs to limit its interference below a threshold. We refer to
this kind of constraint as an interference power constraint (IPC).

In this paper we consider the problem of finding the secrecy capacity-achieving
input covariance for Gaussian MIMO wiretap channels with joint SPC,
PAPC, and IPC. As mentioned above, analytical solutions to the general
Gaussian MIMO WTC have not been reported, and thus efficient numerical
methods are desired. To this end we note that the solution proposed
in \cite{Li2013} suffers a slow convergence rate which is inherent
in subgradient methods. In addition, this method can only yield a
locally optimal solution. On the contrary, the barrier method presented
in \cite{lyoka2015minmaxEQ}, being a Newton-type method, converges
very fast but its per-iteration complexity increases rapidly with
the problem size.\textcolor{black}{{} Thus, our motivation is to develop
more efficient numerical methods to solve the secrecy capacity problem
of the MIMO WTC. To this end we propose two algorithms which can overcome
the shortcomings of these existing solutions. Although, new research
directions in the study of the MIMO WTC are not the scope of this
paper, the proposed numerical methods are still of significant importance
since they will help find the capacity and optimal signaling faster,
which is useful to deal with time-varying channels. For example, in
a closed-loop system, we need to solve the secrecy problem whenever
the channels are updated. In this regard, the secrecy optimization
problem is an on-line task. As a result, our proposed fast converging
algorithms are always desired, and thus, will certainly create an
impact. Our main contributions are as follows:}
\begin{itemize}
\item For the \emph{degraded} Gaussian MIMO WTC, we present a convex reformulation
for finding the optimal signaling and the secrecy capacity. We remark
that in this case, the secrecy capacity problem is in fact convex
but is expressed in a non-convex form. To our knowledge, no convex
reformulation has been reported fo\textcolor{black}{r a general set
up as considered in this paper. The convex reformulation allows us
to solve the secrecy capacity problem using off-the-shelf convex solvers
for different type}s of transmit power constraints for which analytical
solutions are impossible.
\item For the general Gaussian MIMO WTC, we apply an accelerated DCA \cite{NhatPhan2017}
to solve the secrecy capacity problem, which requires solving a sequence
of convex subproblems. To solve these subproblems, we customize the
CoMirror algorithm introduced in \cite{Beck2010} to achieve an iterative
method where each iteration is solved in closed-form. The numerical
results demonstrate that the accelerated DCA converges very much quicker,
compared to a known solution that is based on AO.
\item For the general Gaussian MIMO WTC, we also propose an efficient iterative
method to calculate the secrecy capacity, which is based on the equivalent
concave-convex reformulation of the secrecy capacity problem. We refer
to this proposed method as the partial best response algorithm (PBRA).
The idea of PBRA is to find a saddle point of the concave-convex game,
by optimizing one variable while the other is held fixed. The novelty
of the PBRA is the use of a proper approximation of the saddle-point
objective to achieve monotonic convergence to a saddle point. Also,
each iteration of the PBRA can be solved efficiently.{}
\end{itemize}
\textit{Notation:} We use bold uppercase and lowercase letters to
denote matrices and vectors, respectively. $\mathbb{C}^{M\times N}$
denotes the space of $M\times N$ complex matrices. To lighten the
notation, $\mathbf{I}$ and $\mathbf{0}$ define identity and zero
matrices respectively, of which the size can be easily inferred from
the context. $||\cdot||_{F}$ and $||\cdot||_{1}$ denote the Frobenius
and $\ell_{1}$ norm. $\mathbf{H}\herm$ and $\mathbf{H}\trans$ are
Hermitian and ordinary transpose of $\mathbf{H}$, respectively; $\mathbf{H}_{i,j}$
is the $(i,j)$-entry of $\mathbf{H}$; $|\mathbf{H}|$ is the determinant
of $\mathbf{H}$; Furthermore, we denote the expected value of a random
variable by $E\{.\}$, and $[x]_{+}=\max(x,0)$. The $i$th unit vector
(i.e., its $i$th entry is equal to one and all other entries are
zero) is denoted by $\mathbf{e}_{i}$. The notation $\mathbf{A}\succeq(\succ)\mathbf{B}$
means $\mathbf{A}-\mathbf{B}$ is positive semidefinite (definite).
$\nabla_{\mathbf{X}}f$, where $\mathbf{X}\in\mathbb{C}^{M\times N}$,
denotes the complex gradient with respect to $\mathbf{X}^{\ast}$
as defined in \cite{Hjorungens2007matrixAlgebra}. $P_{\mathcal{C}}(\mathbf{x})$
is the Euclidean projection of $\mathbf{x}$ onto the set $\mathcal{C}$.
$\mathcal{CN}(\mathbf{0},\mathbf{A})$ denotes a circularly-symmetric
complex-valued Gaussian random vector with zero mean and covariance
$\mathbf{A}$

\section{System Model}

Consider a MIMO WTC with a transmitter (Alice), a legitimate receiver
(Bob) and an eavesdropper (Eve). Let $N_{t}$, $N_{r}$, and$N_{e}$
denote the number of antennas at Alice, Bob, and Eve, respectively.
The signals received at Bob and Eve can be expressed as \begin{subequations}
\begin{align}
\mathbf{y}_{b} & =\mathbf{H}_{b}\mathbf{x}+\mathbf{z}_{r}\\
\mathbf{y}_{e} & =\mathbf{H}_{e}\mathbf{x}+\mathbf{z}_{e}
\end{align}
\end{subequations} respectively. In the equation above, $\mathbf{x}$
is the confidential signal that Alice wishes to transmit to Bob, while
keeping it secret from Eve; $\mathbf{H}_{b}\in\mathbb{C}^{N_{r}\times N_{t}}$
and $\mathbf{H}_{e}\in\mathbb{C}^{N_{e}\times N_{t}}$ are the complex
channel matrix between Alice and Bob, and between Alice a and Eve.
$\mathbf{z}_{r}\in\mathbb{C}^{N_{r}\times1}\sim\mathcal{CN}(\mathbf{0},\mathbf{I})$
and $\mathbf{z}_{e}\in\mathbb{C}^{N_{e}\times1}\sim\mathcal{CN}(\mathbf{0},\mathbf{I})$
are additive white Gaussian noise at the legitimate receiver and at
the eavesdropper respectively.\footnote{Note that, for ease of mathematical description we have assumed the
noise at both the legitimate receiver and the eavesdropper and normalized
$\mathbf{H}_{b}$ and $\mathbf{H}_{e}$ to the true noise power and
thus the normalized noise power is unity.} In this paper $\mathbf{H}_{b}$ and $\mathbf{H}_{e}$ are assumed
to be quasi-static and perfectly known at all nodes. For a given input
covariance matrix $\mathbf{X}=E\{\mathbf{x}\mathbf{x}\herm\}\succeq\mathbf{0}$,
where $E\{\cdot\}$ is the statistical expectation, the maximum secrecy
rate (in nat/s/Hz) between Alice and Bob is given by \cite{Oggier2011SecCapEq}
\begin{equation}
C_{s}(\mathbf{X})=\left[\underbrace{\ln|\mathbf{I}+\mathbf{H}_{b}\mathbf{X}\mathbf{H}_{b}\herm|}_{f_{b}(\mathbf{X})}-\underbrace{\ln|\mathbf{I}+\mathbf{H}_{e}\mathbf{X}\mathbf{H}_{e}\herm|}_{f_{e}(\mathbf{X})}\right]_{+}.\label{SecrecyCapacityEqn}
\end{equation}
In this paper we are interested in the secrecy capacity of MIMO WTC
subject to some constraints on the transmit covariance, which is mathematically
stated as 
\begin{equation}
\boxed{C_{s}\triangleq\underset{\mathbf{X}\in\mathcal{X}}{\max}\ C_{s}(\mathbf{X})}\label{eq:secrecycapacity:org}
\end{equation}
where $\mathcal{X}$ is determined by the transmit power constraints
of interest. Some typical examples of $\mathcal{X}$ are given below.
\begin{itemize}
\item The SPC:
\begin{equation}
\mathcal{X}_{\textrm{spc}}=\left\{ \mathbf{X}\succeq\mathbf{0}\ |\ \tr(\mathbf{X})\leq P_{0}\right\} \label{eq:SPC:only}
\end{equation}
where $P_{0}$ is the total power budget. This setting is fundamental
to MIMO WTC and its secrecy capacity has been extensively studied
\cite{MIMOME_WTC,Oggier2011SecCapEq,Sec_MIMO_SPC_3,Loyka2016,ThangNguyen2020}.
In this case the SPC can be set to $\tr(\mathbf{X})=P_{0}$ without
loss of optimality.
\item The PAPC:
\begin{equation}
\mathcal{X}_{\textrm{papc}}=\left\{ \mathbf{X}\succeq\mathbf{0}\ |\ [\mathbf{X}]_{i,i}\leq P_{i},i=1,2,\dots,N_{t}\right\} \label{eq:PAPC}
\end{equation}
 where $P_{i}$ is the maximum allowed power for the $i$-th antenna.
It very often that joint SPC and PAPC (i.e. $\mathcal{X}=\mathcal{X}_{\textrm{spc}}\cap\mathcal{X}_{\textrm{papc}})$
is considered in the literature, e.g., in \cite{Li2013,Li2013b} for
MIMO WTC and recently in \cite{Cao2020} for MISO WTC.
\item The interference power constraint (IPC):
\begin{equation}
\mathcal{X}_{\textrm{ipc}}=\left\{ \mathbf{X}\succeq\mathbf{0}\ |\ \tr(\mathbf{W}_{l}\mathbf{X})\leq P_{l},l=1,2,\ldots,N_{p}\right\} \label{eq:IPC}
\end{equation}
 where $\mathbf{W}_{l}=\mathbf{H}_{l}\herm\mathbf{H}_{l}$ and $\mathbf{H}_{l}$
is the channel between Alice and the $l$-th primary receiver, $P_{l}\geq0$
is the corresponding interference threshold, and $N_{p}$ is the number
of primary receivers. It means that the interference energy from Alice
to the $l$-th primary receiver should be limited by a predetermined
threshold. The case for $\mathcal{X}=\mathcal{X}_{\textrm{spc}}\cap\mathcal{X}_{\textrm{ipc}}$was
studied in \cite{Loyka2018,Dong2018c}.
\end{itemize}
We remark that $C_{s}>0$ if and only if $\mathbf{H}_{b}\herm\mathbf{H}_{b}-\mathbf{H}_{e}\herm\mathbf{H}_{e}$
is positive semidefinite or indefinite, i.e. $\mathbf{H}_{b}\herm\mathbf{H}_{b}-\mathbf{H}_{e}\herm\mathbf{H}_{e}$
has at least one positive eigenvalue, which is assumed in the sequel
of the paper. A proof for this can be found in \cite[Appendix A]{Li2009}.
Thus we can remove the max operator in \eqref{eq:secrecycapacity:org}
onward without loss of optimality. Further, let $\mathbf{X}^{\ast}$
be the optimal input covariance matrix of \eqref{eq:secrecycapacity:org}.
Then it was proved in \cite[Corollary 1]{MIMOME_WTC} that the secrecy
capacity can be achieved by a wiretap coding scheme the follows a
circularly-symmetric complex-valued Gaussian random vector with zero
mean and covariance $\mathbf{X}^{\ast}$. Moreover, for a given $\mathbf{X}^{\ast}$,
a way to construct a Gaussian wiretap code that achieves the secrecy
capacity was presented in \cite{Khina2018:MIMOWTC:decomposed}. The
idea is to apply the generalized singular value decomposition to decompose
the MIMO wiretap channel into several parallel eigen-subchannels.
Then, to achieve the secrecy capacity, Gaussian wiretap codebooks
are sent along the subchannels where the gains to Bob are larger than
those to Eve. We refer the interested readers to \cite{Khina2018:MIMOWTC:decomposed}
for further details. One may argue that considering capacity achieving
schemes as done in this paper is not of practical importance since
discrete modulation schemes and coding rates are used in practice.
However, we note that solving \eqref{eq:secrecycapacity:org} is still
practically meaningful.\textcolor{black}{{} Firstly, it can give an
upper bound on what we can achieve in terms of secrecy rate. }{Secondly,
the obtained optimal covariance matrix can be useful to construct
wiretap codes with practical finite-alphabet input that can achieve
a secrecy rate close to the secrecy capacity up to an SNR threshold
\cite{Bashar2012FiniteAlphabet}.}

\section{\textcolor{black}{Convex Reformulations for MIMO Wiretap Channels}}

\subsection{The Degraded Case}

In general $C_{s}(\mathbf{X})$ is non-convex with respect to $\mathbf{X}$,
and thus, finding optimal signaling for MIMO WTC is difficult. However,
if the channel is degraded (i.e. $\mathbf{H}_{b}\herm\mathbf{H}_{b}\succeq\mathbf{H}_{e}\herm\mathbf{H}_{e}$)
then $C_{s}(\mathbf{X})$ becomes concave (i.e. problem \eqref{eq:secrecycapacity:org}
is convex) \cite{Oggier2011SecCapEq}, and thus, efficient algorithms
for solving \eqref{eq:secrecycapacity:org} are possible in principle.
More specifically, analytical solutions have been reported for degraded
MIMO WTC under some specific power constraints. For example, full-rank
solutions via water-filling like algorithm for the SPC only was presented
in \cite{Sec_MIMO_SPC_3}. Moreover, when the transmit power is sufficient
large, closed-form for optimal signaling is possible \cite{Loyka2016}.
When $\mathcal{X}$ is either SPC only or PAPC only, closed-form solutions
are presented in \cite{Cao2020} for MISO WTC. When $\mathcal{X}$
stands for the joint SPC and PAPC, numerical solutions based on alternating
optimization are proposed in \cite{Li2013} for MIMO WTC.

Regarding numerical algorithms for finding optimal signaling for degraded
MIMO WTC, we note that off-the-shelf convex solvers cannot be used
to solve \eqref{eq:secrecycapacity:org} directly despite its convexity,
since it is not expressed in a standard convex form. To overcome this
issue, one could customize standard algorithms for convex optimization
such as interior-point methods or gradient based methods to solve
\eqref{eq:secrecycapacity:org}, which was done e.g. in \cite{lyoka2015minmaxEQ}.In
this section, we equivalently reformulate \eqref{eq:secrecycapacity:org}
as a standard convex problem for degraded MIMO WTC. As a result, we
can avail of powerful modern convex solvers to compute the optimal
signaling. In this regard, the following lemma is in order.
\begin{lem}
\label{lem:convexreform}Let $\boldsymbol{\Delta}=\mathbf{H}_{b}\herm\mathbf{H}_{b}-\mathbf{H}_{e}\herm\mathbf{H}_{e}\succeq\mathbf{0}$.
Then problem \eqref{eq:secrecycapacity:org} is equivalent to the
following convex problem \begin{subequations}\label{eq:secrecyeq}
\begin{align}
\underset{\mathbf{X}\succeq\mathbf{0},\mathbf{Y}\succeq\mathbf{0}}{\mathrm{maximize}} & \ \ln|\mathbf{Y}|\\
\st & \ \begin{bmatrix}\mathbf{I}+\boldsymbol{\Delta}^{1/2}\mathbf{X}\boldsymbol{\Delta}^{1/2}-\mathbf{Y} & \boldsymbol{\Delta}^{1/2}\mathbf{X}\mathbf{H}_{e}\herm\\
\mathbf{H}_{e}\mathbf{X}\boldsymbol{\Delta}^{1/2} & \mathbf{I}+\mathbf{H}_{e}\mathbf{X}\mathbf{H}_{e}\herm
\end{bmatrix}\succeq\mathbf{0}\label{eq:Slemma}\\
 & \ \mathbf{X}\in\mathcal{X}
\end{align}
where $\mathcal{X}$ denotes any convex set of transmit covariance
constraints including those in \eqref{eq:SPC:only}, \eqref{eq:PAPC},
\eqref{eq:IPC}, or any combination thereof. \end{subequations} 
\end{lem}
\begin{IEEEproof}
\textcolor{black}{Please refer to Appendix \ref{lem:convexreform}.}
\end{IEEEproof}
\textcolor{black}{We note that a similar convex reformulation was
also presented for degraded channels in \cite{Shi2015CovReform} for
the SPC and Alice also sends some power to an energy harvester which
acts as an eavesdropper. The proof in \cite{Shi2015CovReform} is
quite involved. In our paper, we only require the feasible set of
the secrecy capacity problem to be convex. Thus, it can deal with
any linear transmit covariance matrix constraints, including energy
harvesting threshold constraint as a special case. We remark that
our proof for the convex reformulation is based on the epigraph form
of \eqref{eq:secrecycapacity:org} and is much more elegant. We further
remark that \eqref{eq:secrecyeq} can be converted into a standard
semidefinite program, which is done automatically by modeling tools
for convex optimization such as CVX \cite{CVX} and YALMIP \cite{YALMIP}.
The interested reader is referred to \cite[p. 149]{Bental:2001} for
further details. To conclude this section we note that modern off-the-shelf
solvers such as MOSEK \cite{mosek} can solve \eqref{eq:secrecyeq}
very fast when $N_{t}$ is not too large.}

\subsection{\textcolor{black}{The General Case}}

For nondegraded MIMO WTC, problem \eqref{eq:secrecycapacity:org}
is a non-convex program in general, and thus, finding optimal signaling
is difficult. In such cases, a convex-concave reformulation of \eqref{eq:secrecycapacity:org}
is particularly useful. Specifically, based on the collective results
in \cite{Bustin2009,Oggier2011SecCapEq,MIMOME_WTC,lyoka2015minmaxEQ,Dong2018c},
the secrecy capacity of general MIMO WTC in \eqref{eq:secrecycapacity:org}
can be equivalently expressed in the form of a minimax optimization
problem as 
\begin{equation}
\boxed{C_{s}=\underset{\mathbf{K}\in\mathcal{K}}{\min}\ \underset{\mathbf{X}\in\mathcal{X}}{\max}\ f(\mathbf{K},\mathbf{X})\triangleq\log\frac{|\mathbf{I}+\mathbf{K}^{-1}\mathbf{H}\mathbf{X}\mathbf{H}\herm|}{|\mathbf{I}+\mathbf{H}_{e}\mathbf{X}\mathbf{H}_{e}\herm|}}\label{eq:minimax}
\end{equation}
where $\mathbf{H}=[\mathbf{H}_{b}\trans,\mathbf{H}_{e}\trans]\trans\in\mathbb{C}^{(N_{r}+N_{e})\times N_{t}}$
is the extended channel, $\mathcal{X}$ stands for the transmit power
constraints including SPC, PAPC and IPC, or any combination thereof,
and $\mathcal{K}$ is defined as
\begin{equation}
\mathcal{K}=\left\{ \mathbf{K}\left|\mathbf{K}=\small\left[\begin{array}{cc}
\mathbf{I} & \bar{\mathbf{K}}\\
\bar{\mathbf{K}}\herm & \mathbf{I}
\end{array}\right];\bar{\mathbf{K}}\in\mathbb{C}^{N_{r}\times N_{e}};\mathbf{K}\succeq\mathbf{0}\right.\right\} .
\end{equation}
The set $\mathcal{K}$ deserves further explanations. In fact, $\mathbf{K}$
in \eqref{eq:minimax} is the covariance matrix of the following composite
noise:
\begin{equation}
\mathbf{z}=\begin{bmatrix}\mathbf{z}_{r}\\
\mathbf{z}_{e}
\end{bmatrix}
\end{equation}
which is obtained by assuming that Bob knows both $\mathbf{H}_{b}$
and $\mathbf{H}_{e}$ \cite{MIMOME_WTC,Oggier2011SecCapEq}. As a
result, $\mathbf{K}$ is defined as
\[
\mathbf{K}=\left[\begin{array}{cc}
E\bigl\{\mathbf{z}_{r}\mathbf{z}_{r}\herm\bigr\} & E\bigl\{\mathbf{z}_{r}\mathbf{z}_{e}\herm\bigr\}\\
E\bigl\{\mathbf{z}_{e}\mathbf{z}_{r}\herm\bigr\} & E\bigl\{\mathbf{z}_{e}\mathbf{z}_{e}\herm\bigr\}
\end{array}\right]=\left[\begin{array}{cc}
\mathbf{I} & \bar{\mathbf{K}}\\
\bar{\mathbf{K}}\herm & \mathbf{I}
\end{array}\right],
\]
where $\bar{\mathbf{K}}$ represents the correlation between $\mathbf{z}_{r}$
and $\mathbf{z}_{e}$.

\textcolor{black}{We remark that \eqref{eq:minimax} is true regardless
of the degradedness of the MIMO WTC. The significance of the above
minimax reformulation is two fold. First, computing the secrecy capacity
is always equivalent to finding a saddle point of \eqref{eq:minimax}.}
Second, \eqref{eq:minimax} is more numerically tractable since $f(\mathbf{K},\mathbf{X})$
is convex with respect to $\mathbf{K}$ and is concave with respect
to $\mathbf{X}$. Thus, \eqref{eq:minimax} is also widely known as
a convex-concave problem. Let $(\mathbf{X}^{\ast},\mathbf{K}^{\ast})$
be the saddle point of \eqref{eq:minimax} which always exists since
$\mathcal{X}$ and $\mathcal{K}$ are convex and compact. Then the
following inequality holds
\begin{equation}
f(\mathbf{X},\mathbf{K}^{\ast})\leq C_{s}=f(\mathbf{X}^{\ast},\mathbf{K}^{\ast})\leq f(\mathbf{X}^{\ast},\mathbf{K}).
\end{equation}
Further $f(\mathbf{K},\mathbf{X})$ is an upper bound of $C_{s}(\mathbf{X})$,
i.e. $f(\mathbf{K},\mathbf{X})\geq C_{s}(\mathbf{X})$ for any feasible
$(\mathbf{K},\mathbf{X})$. However, it is worth noting that, while
the\textcolor{black}{{} equality} $\underset{\mathbf{K}\in\mathcal{K}}{\min}\ \underset{\mathbf{X}\in\mathcal{X}}{\max}\ f(\mathbf{K},\mathbf{X})=\underset{\mathbf{X}\in\mathcal{X}}{\max}\ C_{s}(\mathbf{X})$
is always true, a saddle point $(\mathbf{X}^{\ast},\mathbf{K}^{\ast})$
to \eqref{eq:minimax} is not necessarily an optimal solution to \eqref{eq:secrecycapacity:org}
in general. More precisely, it is possible that $C_{s}=f(\mathbf{X}^{\ast},\mathbf{K}^{\ast})>C_{s}(\mathbf{X}^{\ast})$
for some cases, especially when \eqref{eq:minimax} has multiple saddle
points. For example, consider the following real-valued channel matrices
for simplicity:
\begin{equation}
\mathbf{H}_{b}=\begin{bmatrix}-0.4176 & 1.4224\\
-1.4963 & -2.0426
\end{bmatrix};\mathbf{H}_{e}=\begin{bmatrix}0.6726 & 1.4335\\
1.7762 & -0.3694
\end{bmatrix}.
\end{equation}
Note that the resulting MIMO WTC is nondegraded and thus convex reformulation
presented in the preceding subsection is not applicable. For the joint
SPC and PAPC given in \eqref{eq:PAPC} with $P_{0}=10$, $P_{1}=P_{2}=6$,
solving \eqref{eq:minimax} (using the minimax barrier method in \cite{lyoka2015minmaxEQ}
or Algorithm \ref{alg:PBRA} presented shortly) yields 
\begin{equation}
\mathbf{X}^{\ast}=\begin{bmatrix}1.7305 & 1.2198\\
1.2198 & 5.9985
\end{bmatrix}\ \textrm{and}\ C_{s}=1.0420,
\end{equation}
but $C_{s}(\mathbf{X}^{\ast})=0.3409<C_{s}$. However, if the SPC
is active, i.e. $\tr(\mathbf{X}^{\ast})=P_{0}$, then $\mathbf{X}^{\ast}$
is also a maximizer of \eqref{eq:secrecycapacity:org}. The above
example implies that numerical algorithms for solving both \eqref{eq:secrecycapacity:org}
and \eqref{eq:minimax} are desired.

To motivate the efficient numerical methods proposed in the subsequent
sections we note tha\textcolor{black}{t existing numerical solutions
for solving \eqref{eq:secrecycapacity:org} for general MIMO WTC can
be generally classified into two ways. The first one is based on local
optimization approaches to solving \eqref{eq:secrecycapacity:org}
directly with the hope that they can also yield an optimal solution
by a good initialization \cite{Li2013,Li2013b,Steinwandt2014}. The
drawback of such methods is that the achieved covariance matrix is
not guaranteed to be the optimal signaling. The second way is based
on finding a saddle point of a convex-concave reformulation of \eqref{eq:secrecycapacity:org}
\cite{lyoka2015minmaxEQ}. However, as explained by the example above,
such a method always gives the secrecy capacity but not necessarily
the optimal signaling. More explicitly, if we construct a Gaussian
wiretap code based on the obtained saddle point of \eqref{eq:minimax},
then the achievable secrecy rate can be strictly smaller than the
secrecy capacity.}

\subsection{\textcolor{black}{A Suboptimal Method}}

For comparison purpose we briefly describe a suboptimal method that
can efficiently compute an achievable secrecy rate. In particular,
this method is obtained by forcing $\mathbf{H}_{e}\mathbf{X}\mathbf{H}_{e}\herm=0$.
Note that we can rewrite $\mathbf{X}=\mathbf{U}\mathbf{U}\herm$ for
some $\mathbf{U}$. Thus the constraint $\mathbf{H}_{e}\mathbf{X}\mathbf{H}_{e}\herm=\mathbf{0}$
is equivalent to $\mathbf{H}_{e}\mathbf{U}=\mathbf{0}$, which means
$\mathbf{\mathbf{U}}$ should belong to the null space of $\mathbf{H}_{e}$.
Let $\mathbf{V}$ be a basis of the null space of $\mathbf{H}_{e}$
which is nonempty when $N_{t}>N_{e}$. Then we can write $\mathbf{X}=\mathbf{V}\mathbf{T}\mathbf{V}\herm$,
where $\mathbf{T}\succeq\mathbf{0}$ is the solution to the following
optimization problem \begin{subequations}
\begin{align}
\underset{\mathbf{T}\succeq\mathbf{0}}{\maximize} & \;\ln|\mathbf{I}+\mathbf{H}_{b}\mathbf{V}\mathbf{T}\mathbf{V}\herm\mathbf{H}_{b}\herm|\\
\st & \mathbf{\;}\mathbf{V}\mathbf{T}\mathbf{V}\herm\in\mathcal{X}.
\end{align}
\end{subequations} In the remainder of the paper we refer to this
suboptimal method as the zero-forcing (ZF) method since the idea in
fact comes from the zero-forcing method for downlink multiuser MIMO
\cite{Spencer2004}.

\section{Accelerated DCA Method for Solving \eqref{eq:secrecycapacity:org}}

\subsection{Algorithm Description}

As mentioned above, since the equivalent convex-concave formulation
is not always useful to find the optimal signaling of the general
MIMO WTC, one still needs to solve \eqref{eq:secrecycapacity:org}
directly. In \cite{Li2013,Li2013b}, an AO method was introduced to
solve \eqref{eq:secrecycapacity:org}. Here we propose a simple but
efficient method derived based on the obvious observation that $C_{s}(\mathbf{X})$
is a DC function. Note that $f_{b}(\mathbf{X})$ and $f_{e}(\mathbf{X})$
are indeed concave functions \cite[Section 3.1]{boyd_vandenberghe_2004}
and $C_{s}(\mathbf{X})$ can be rewritten as $C_{s}(\mathbf{X})=-f_{e}(\mathbf{X})-(-f_{b}(\mathbf{X}))$
which is a DC function. This naturally motivates the use of DCA to
solve \eqref{eq:secrecycapacity:org}.\emph{ }In this regard maximizing
a concave function is a convex problem, and thus, the term $-f_{e}(\mathbf{X})$
is considered as the non-convex term. Thus, the main idea of the conventional
DCA is to linearize the non-convex term of the problem, which is $-f_{e}(\mathbf{X})$
in our case, at a given operating point and solve the approximate
convex subproblem. This process is repeated until some stopping criterion
is met.

In this paper we consider an accelerated version of DCA (ADCA) presented
in \cite{NhatPhan2017}. The idea is that from the current and previous
iterates denoted by $\mathbf{X}_{n}$ and $\mathbf{X}_{n-1}$ respectively,
we compute an extrapolated point $\mathbf{Z}_{n}$ using the Nesterov\textquoteright s
acceleration technique: $\mathbf{X}_{n}+(t_{k}-1)/t_{k+1}\bigl(\mathbf{X}_{n}-\mathbf{X}_{n-1}\bigr)$.
Since $C_{s}(\mathbf{X})$ is possibly non-convex for a general MIMO
WTC, $\mathbf{Z}_{n}$ can be a bad extrapolation and a monitor is
required. Specifically, if $\mathbf{Z}_{n}$ is better than \emph{one
of the last} $q$ iterates, then $\mathbf{Z}_{n}$ is considered a
good extrapolation and thus will be used instead of $\mathbf{X}_{n}$
to generate the next iterate. Thus, the ADCA is generally \emph{non-monotone}.
The algorithmic description of ADCA for solving \eqref{eq:secrecycapacity:org}
is outlined in Algorithm \ref{alg:ADCA}. Note that the subproblem
in \eqref{eq:DCA:subprob} is achieved by linearizing $f_{e}(\mathbf{X})$
around $\mathbf{V}_{n}$ and omitting the associated constants that
do not affect the optimization. In Algorithm \ref{alg:ADCA}, $q$
is any non-negative integer and $\gamma_{n}$ is the minimum of the
secrecy rate of the last $q$ iterates. We remark that the case when
$q=0$ reduces to the conventional DCA, which is exactly the same
as the AO method in \cite{Li2013}.
\begin{algorithm}[tbh]
\caption{ADCA for solving \eqref{eq:secrecycapacity:org}}

\label{alg:ADCA}

\begin{algorithmic}[1]

\STATE Initialization: $\mathbf{V}_{0}=\mathbf{X}_{0}\in\mathcal{X}$,
$t=\frac{1+\sqrt{5}}{2}$, $q$: integer.

\FOR{ $n=1,2,\dots$}

\STATE Update:
\begin{equation}
\mathbf{X}_{n}=\underset{\mathbf{X}\in\mathcal{X}}{\arg\max}\ \underbrace{f_{b}(\mathbf{X})-\tr\bigl(\nabla f_{e}(\mathbf{V}_{n-1})\mathbf{X}\bigr)}_{\bar{f}(\mathbf{X};\mathbf{V}_{n-1})}\label{eq:DCA:subprob}
\end{equation}
where $\nabla f_{e}(\mathbf{X})=\mathbf{H}_{e}\herm\bigl(\mathbf{I}+\mathbf{H}_{e}\mathbf{X}\mathbf{H}_{e}\herm\bigr)^{-1}\mathbf{H}_{e}$ 

\STATE $t_{n+1}=\frac{1+\sqrt{1+4t_{n}^{2}}}{2}$

\STATE $\mathbf{Z}_{n}=\mathbf{X}_{n}+\frac{t_{n}-1}{t_{n+1}}\bigl(\mathbf{X}_{n}-\mathbf{X}_{n-1}\bigr)$

\STATE $\gamma_{n}=\min\bigl(C_{s}(\mathbf{X}_{n}),C_{s}(\mathbf{X}_{n-1}),\ldots,C(\mathbf{X}_{[n-q]_{+}})\bigr)$\label{alg:ADCA:gamma}

\STATE $\mathbf{V}_{n}=\begin{cases}
\mathbf{Z}_{n} & \textrm{if}\ C_{s}(\mathbf{Z}_{n})\geq\gamma_{n}\\
\mathbf{X}_{n} & \textrm{otherwise}
\end{cases}$\label{alg:ADCA:updateV}

\ENDFOR

\STATE Output: $\mathbf{X}_{n}$

\end{algorithmic}
\end{algorithm}
\textcolor{black}{Before proceeding further we also note that Algorithm
\ref{alg:ADCA} in our paper is not a traditional first-order Taylor
method. In particular, we apply an extrapolated point which is numerically
shown to improve the convergence rate.}

\subsection{\textcolor{black}{Convergence Analysis}}

The convergence analysis of the ADCA is studied in \cite{NhatPhan2017}
where the involved functions are assumed to be strongly convex. In
the considered problem, this assumption does not hold for $f_{b}(\mathbf{X})$
and $f_{e}(\mathbf{X})$ in general. A weaker convergence is stated
in the following lemma.
\begin{lem}
\label{lem:ADCA:convergence}Let $\{\gamma_{n}\}$ be the sequence
generated by Step \ref{alg:ADCA:gamma} of Algorithm \ref{alg:ADCA}.
Then it holds that $\gamma_{n+q}\geq\gamma_{n-1}$. If the objective
$\bar{f}(\mathbf{X};\mathbf{V}_{n-1})$ is strictly concave. e.g.
when $\mathbf{H}_{b}\herm\mathbf{H}_{b}$ is invertible, then, every
limit points of Algorithm \ref{alg:ADCA} is a critical point of \ref{eq:secrecycapacity:org}.
\end{lem}
\begin{IEEEproof}
Please refer to Appendix \ref{subsec:proof:ADCA:convergence}.
\end{IEEEproof}

\subsection{Solving the Subproblem for joint SPC and PAPC: CoMirror Algorithm}

To implement Algorithm \ref{alg:ADCA}, we need to solve \eqref{eq:DCA:subprob}
efficiently. We remark that for the case of SPC only, waterfilling-like
solution to \eqref{eq:DCA:subprob} is possible. We skip the details
here for the sake of brevity. Thus we focus on the joint SPC and PAPC
case, i.e. $\mathcal{X}=\mathcal{X}_{\textrm{spc}}\cap\mathcal{X}_{\textrm{papc}}$
where $\mathcal{X}_{\textrm{spc}}$ and $\mathcal{X}_{\textrm{papc}}$
are defined in \eqref{eq:SPC:only} and \eqref{eq:PAPC}, respectively.
Since \eqref{eq:DCA:subprob} is a convex program, convex solvers
can be used to solve it. However, the incurred complexity (including
the memory requirement) is very high when $N_{t}$ is large, which
is the case for massive MIMO. Our goal in this section is to derive
a more efficient method for solving \eqref{eq:DCA:subprob}. To this
end we note that the spectrahedron $\mathcal{X}_{\textrm{spc}}$ is
simple in the sense that the projection onto it can be computed efficiently
as shall be seen shortly. To exploit this fact, we resort to the CoMirror
algorithm presented in \cite{Beck2010} to solve \eqref{eq:DCA:subprob}. 

To simplify the notation we will ignore $\mathbf{V}_{n-1}$ and write
$\bar{f}(\mathbf{X})$ instead of $\bar{f}(\mathbf{X};\mathbf{V}_{n-1})$
onward. Let $g_{i}(\mathbf{X})=[\mathbf{X}]_{i,i}-P_{i}$. Then \eqref{eq:DCA:subprob}
is equivalent to\begin{subequations}\label{eq:DCA:subprob:comp}
\begin{align}
\underset{\mathbf{X}\in\mathcal{X}_{\textrm{spc}}}{\maximize} & \quad\bar{f}(\mathbf{X})\\
\st & \quad g_{i}(\mathbf{X})\leq0,i=1,2,\ldots,N_{t}\label{eq:PAPC:rewrite}
\end{align}
\end{subequations} The operation of the CoMirror algorithm is as
follows.\footnote{Specifically, we particularize the CoMirror algorithm in \cite{Beck2010}
for the Euclidean setting and adapt the description to fit the maximization
context.} For a given iterate $\mathbf{X}^{k}$, if the constraint \eqref{eq:PAPC:rewrite}
is satisfied, then we move along the direction $\nabla\bar{f}_{t}(\mathbf{X})$
with a step size $\eta_{k}$ to maximize the objective, generating
the next iterate. If \eqref{eq:PAPC:rewrite} is violated, set $m=\underset{1\leq i\leq N_{t}}{\arg\max}\;g_{i}(\mathbf{X})$
and move along $-\nabla g_{m}(\mathbf{X})=-\diag(\mathbf{e}_{m})$
to reduce $g_{m}(\mathbf{X})$, hoping to achieve a feasible solution
in the next iteration. The CoMirror algorithm for solving \eqref{eq:DCA:subprob}
is summarized in Algorithm \ref{alg:CoMirror}. The convergence of
Algorithm \ref{alg:CoMirror} and other relevant discussions are provided
in Appendix \ref{sec:convergence:comirror}.
\begin{algorithm}[tbh]
\caption{CoMirror algorithm for solving \eqref{eq:DCA:subprob}}

\label{alg:CoMirror}

\begin{algorithmic}[1]

\STATE Initialization: $\mathbf{X}^{0}\in\mathcal{X}_{\textrm{spc}}$;
$\Omega=\frac{1}{\sqrt{2}}\underset{\mathbf{X}\in\mathcal{X}_{\textrm{spc}}}{\max}\ \bigl\Vert\mathbf{X}-\mathbf{X}^{0}\bigr\Vert$;

\FOR{ $k=1,2\dots$}

\STATE $\mathbf{X}^{k}=P_{\mathcal{X}_{\textrm{spc}}}\left(\mathbf{X}^{k-1}+\eta_{k}\mathbf{E}_{k-1}\right)$
where 
\[
\mathbf{E}_{k-1}=\begin{cases}
\nabla\bar{f}(\mathbf{X}_{k-1}) & \underset{i=1,2,\dots,N_{t}}{\max}\{g_{i}(\mathbf{X}_{k-1})\}\leq0\\
-\diag(\mathbf{e}_{m}) & \textrm{otherwise}
\end{cases}
\]
and 
\[
\eta_{k}=\frac{\Omega}{\bigl\Vert\mathbf{E}_{k-1}\bigr\Vert\sqrt{k}}
\]

\ENDFOR

\STATE Output: $\mathbf{X}_{k}$

\end{algorithmic}
\end{algorithm}

The following remarks are in order regarding the implementation of
Algorithm \ref{alg:CoMirror}. First, in this paper we use the complex
gradient of $\bar{f}(\mathbf{X}_{k-1})$ defined in \cite{Hjorungens2007matrixAlgebra}
and thus $\nabla\bar{f}(\mathbf{X}_{k-1})$ is given by 
\[
\nabla\bar{f}(\mathbf{X}_{k-1})=\mathbf{H}_{b}\herm(\mathbf{I}+\mathbf{H}_{b}\mathbf{X}_{k-1}\mathbf{H}_{b}\herm)^{-1}\mathbf{H}_{b}-\boldsymbol{\Gamma}_{n-1}.
\]
Second, for a given point $\bar{\mathbf{X}}$, the projection $P_{\mathcal{X}_{\textrm{spc}}}(\bar{\mathbf{X}})$
is mathematically stated as 
\begin{align}
\underset{\mathbf{X}\succeq\mathbf{0}}{\maximize} & \quad\bigl\Vert\mathbf{X}-\bar{\mathbf{X}}\bigr\Vert_{F}^{2}\\
 & \quad\tr(\mathbf{X})\leq P_{0}.
\end{align}
Let $\bar{\mathbf{X}}=\mathbf{U}\diag(\bar{\boldsymbol{\sigma}})\mathbf{U}\herm$
be the eigenvalue decomposition of $\bar{\mathbf{X}}$ and $\bar{\boldsymbol{\sigma}}\in\mathbb{R}^{N_{t}}$
is the vector of the eigenvalues of $\bar{\mathbf{X}}$. Further,
let $\bar{\boldsymbol{\sigma}}_{+}=\max(\bar{\boldsymbol{\sigma}},0)$.
Then the solution to the above problem is given by 
\begin{equation}
\mathbf{X}=\mathbf{U}\diag(P_{\Delta}(\bar{\boldsymbol{\sigma}}_{+}))\mathbf{U}\herm
\end{equation}
where $\Delta$ denotes the full simplex is defined as 
\begin{equation}
\Delta=\{\mathbf{t}\in\mathbb{R}^{N_{t}}|\sum\nolimits _{i=0}^{N_{t}}t_{i}\leq P_{0},t_{i}\geq0,\forall i=1,2,\ldots,N_{t}\}
\end{equation}
and $P_{\Delta}(\bar{\boldsymbol{\sigma}}_{+})$ is given by
\begin{equation}
P_{\Delta}(\bar{\boldsymbol{\sigma}}_{+})=\begin{cases}
\bar{\boldsymbol{\sigma}}_{+} & \textrm{if }\mathbf{1}\trans\bar{\boldsymbol{\sigma}}_{+}\leq P_{0}\\
\bar{\boldsymbol{\sigma}}_{+}-\tau & \textrm{otherwise}
\end{cases}
\end{equation}
where $\tau$ is the unique number such that $\sum_{i=1}^{N_{t}}\max(\bigl[\bar{\boldsymbol{\sigma}}_{+}\bigr]_{i}-\tau,0)=P_{0}$.
Several efficient methods to compute $\tau$ are presented in \cite{Condat2016}.
Overall the per-iteration complexity of Algorithm \ref{alg:CoMirror}
is dominated by that of the EVD of an $N_{t}\times N_{t}$ Hermitian
matrix, which is similar to that of the subgradient method proposed
in \cite{Li2013}. However, we demonstrate in Section \ref{sec:Numerical-Results}
that Algorithm \ref{alg:CoMirror} requires much fewer iterations
to converge.
\begin{rem}
To conclude this section we remark that the above proposed algorithms
can be easily modified to find the optimal signaling of MIMO WTC subject
to joint SPC and IPC, i.e. when $\mathcal{X}=\mathcal{X}_{\textrm{spc}}\cap\mathcal{X}_{\textrm{ipc}}$.
The details are skipped for the sake of brevity.
\end{rem}

\section{Partial Best Response Method for Solving \eqref{eq:minimax}}

We now turn our focus on solving the equivalent minimax reformulation
of the secrecy capacity problem given in \eqref{eq:minimax}. We can
view \eqref{eq:minimax} as a concave-convex game. In a pure best
response algorithm, $\mathbf{X}$ and $\mathbf{K}$ individually maximize
their own goal, given the response of the other. However, since $\mathbf{X}$
and $\mathbf{K}$ are coupled, the best response algorithm (i.e. alternatively
optimizing $\mathbf{X}$ and $\mathbf{K}$) may fail to convergence.
In this paper we propose what so called a partial best response algorithm
(PBRA) which works as follows.

Suppose $\mathbf{K}_{n}$ is available computed at the $n$-th iteration.
Then $\mathbf{X}_{n}$ is found as 
\begin{align}
\mathbf{X}_{n} & =\underset{\mathbf{X}\in\mathcal{X}}{\arg\max}\ f(\mathbf{K}_{n},\mathbf{X})\nonumber \\
 & =\underset{\mathbf{X}\in\mathcal{X}}{\arg\max}\ \log|\mathbf{K}_{n-1}+\mathbf{H}\mathbf{X}\mathbf{H}\herm|-\log|\mathbf{I}+\mathbf{H}_{e}\mathbf{X}\mathbf{H}_{e}\herm|.\label{eq:findX}
\end{align}
In words, $\mathbf{X}_{n}$ is the best response to $\mathbf{K}_{n-1}$
as usual. From the minimax reformulation in \eqref{eq:minimax}, it
is obvious that we need to find the worst noise to achieve the capacity.
The idea of the proposed PBRA is to compute a ``worse noise'' after
each iteration. To this end, we adopt the DCA again where the non-convex
part is linearized. More specifically, given $\mathbf{X}_{n}$, due
to the concavity of the term $\log|\mathbf{K}+\mathbf{H}\mathbf{X}\mathbf{H}\herm|$,
the following inequality holds
\begin{align}
f(\mathbf{K},\mathbf{X}_{n}) & \leq\log|\mathbf{K}_{n-1}+\mathbf{H}\mathbf{X}_{n}\mathbf{H}\herm|+\tr(\boldsymbol{\Psi}_{n}(\mathbf{K}-\mathbf{K}_{n-1}))\nonumber \\
 & \quad-\log(\mathbf{K})-\log|\mathbf{I}+\mathbf{H}_{e}\mathbf{X}_{n}\mathbf{H}_{e}\herm|,\forall\mathbf{K}\in\mathcal{K}.\label{eq:findK:UB}\\
 & \triangleq\bar{f}(\mathbf{K},\mathbf{X}_{n}).
\end{align}
where $\boldsymbol{\Psi}_{n}=(\mathbf{K}_{n-1}+\mathbf{H}\mathbf{X}_{n}\mathbf{H}\herm)^{-1}$.
The inequality is tight when $\mathbf{K}=\mathbf{K}_{n-1}$. Next,
$\mathbf{K}_{n}$ is obtained as
\begin{equation}
\mathbf{K}_{n}=\underset{\mathbf{K}\in\mathcal{K}}{\arg\min}\bar{f}(\mathbf{K},\mathbf{X}_{n})=\underset{\mathbf{K}\in\mathcal{K}}{\arg\min}\tr(\boldsymbol{\Psi}_{n}\mathbf{K})-\log|\mathbf{K}|\label{eq:findK}
\end{equation}
That is to say, $\mathbf{K}_{n}$ is found be the best response to
$\mathbf{X}_{n}$ using\emph{ an upper bound} of the objective. The
proposed solution for finding the secrecy capacity is summarized in
Algorithm \ref{alg:PBRA}. The solutions to the $\mathbf{X}$ and
$\mathbf{K}$ updates are described in the next two subsections.
\begin{algorithm}[tbh]
\caption{PBRA for solving \eqref{eq:minimax}}

\label{alg:PBRA}

\begin{algorithmic}[1]

\STATE Input: $\mathbf{K}_{1}\ensuremath{\in}\mathcal{K}$, $\epsilon_{1}>0$

\FOR{ $n=1,2\dots$}

\STATE Update $\mathbf{X}_{n}$ according to \eqref{eq:findX}

\STATE Update $\mathbf{K}_{n+1}$ according to \eqref{eq:findK}

\ENDFOR

\STATE Output: $\mathbf{X}_{n}$

\end{algorithmic}
\end{algorithm}

\subsection{Efficient Solution for Solving \eqref{eq:findX}}

To implement the proposed PBRA, we need to solve \eqref{eq:findX}.
There are two ways to do this. First, for a given $\mathbf{K}_{n-1}\succ\mathbf{0}$,
\eqref{eq:findX} is equivalent to 
\begin{equation}
\mathbf{X}_{n}=\underset{\mathbf{X}\in\mathcal{X}}{\arg\max}\ \log|\mathbf{I}+\mathbf{K}_{n-1}^{-1/2}\mathbf{H}\mathbf{X}\mathbf{H}\herm\mathbf{K}_{n-1}^{-1/2}|-\log|\mathbf{I}+\mathbf{H}_{e}\mathbf{X}\mathbf{H}_{e}\herm|.
\end{equation}
It is now obvious that the above maximization can be reformulated
as a standard convex problem using Lemma \ref{lem:convexreform} by
simply replacing $\mathbf{H}_{b}$ by $\mathbf{K}_{n-1}^{-1/2}\mathbf{H}$.
The second method is to modify Algorithm \ref{alg:CoMirror} to solve
\eqref{eq:findX}, which can be done straightforwardly. In this regard
the gradient of $f(\mathbf{K}_{n-1},\mathbf{X})$ is given by
\[
\nabla f(\mathbf{K}_{n-1},\mathbf{X})=\mathbf{H}\herm\bigl(\mathbf{K}_{n-1}+\mathbf{H}\mathbf{X}\mathbf{H}\herm\bigr)^{-1}\mathbf{H}-\mathbf{H}_{e}\herm\bigl(\mathbf{I}+\mathbf{H}_{e}\mathbf{X}\mathbf{H}_{e}\herm\bigr)^{-1}\mathbf{H}_{e}.
\]

\subsection{Closed-form solution to \eqref{eq:findK}}

We now show that the $\mathbf{K}$ update admits closed-form solution.
To proceed, we first partition $\boldsymbol{\Psi}_{n}$ into
\begin{equation}
\boldsymbol{\Psi}_{n}=\begin{bmatrix}\boldsymbol{\Psi}_{n,11} & \boldsymbol{\Psi}_{n,12}\\
\boldsymbol{\Psi}_{n,12}^{H} & \boldsymbol{\Psi}_{n,22}
\end{bmatrix}
\end{equation}
where $\boldsymbol{\Psi}_{n,12}\in\mathbb{C}^{n_{R}\times n_{E}}$.
To lighten the notation, we will drop the subscript $n$ in this subsection.
Next let $h(\bar{\mathbf{K}})$ be defined as
\begin{equation}
h(\bar{\mathbf{K}})=\tr\bigl(\boldsymbol{\Psi}_{12}\bar{\mathbf{K}}\herm\bigr)+\tr\bigl(\boldsymbol{\Psi}_{21}\bar{\mathbf{K}}\bigr)-\log\bigl|\mathbf{I}-\bar{\mathbf{K}}\bar{\mathbf{K}}\herm\bigr|.\label{eq:EquivalentKbar}
\end{equation}
Then problem \eqref{eq:findK} is equivalent to the following program\begin{subequations}\label{eq:findKbar}
\begin{align}
\underset{\bar{\mathbf{K}}}{\minimize} & \ h(\bar{\mathbf{K}})\label{eq:minKbar}\\
\st & \ \mathbf{I}-\bar{\mathbf{K}}\bar{\mathbf{K}}\herm\succeq\mathbf{0}.
\end{align}
\end{subequations}The following lemma is in order.
\begin{lem}
\label{lem:optsol:findKbar}Let $\boldsymbol{\Psi}_{12}\boldsymbol{\Psi}_{12}\herm=\mathbf{U}_{\mathbf{\boldsymbol{\Psi}}}\bar{\boldsymbol{\Sigma}}_{\mathbf{\mathbf{\boldsymbol{\Psi}}}}\mathbf{U}_{\mathbf{\boldsymbol{\Psi}}}^{\dagger}$
be the eigenvalue decomposition of $\boldsymbol{\Psi}_{12}\boldsymbol{\Psi}_{12}\herm$
and $\bar{\boldsymbol{\Sigma}}_{\mathbf{\mathbf{\boldsymbol{\Psi}}}}=\diag(\sigma_{\boldsymbol{\Psi}_{1}},\sigma_{\boldsymbol{\Psi}_{2}},\ldots,\sigma_{\boldsymbol{\Psi}_{N_{r}}})$.
Then the optimal solution to \eqref{eq:findKbar} is given by
\begin{equation}
\bar{\mathbf{K}}=-\mathbf{U}_{\boldsymbol{\Psi}}\Xi_{\boldsymbol{\Psi}}\mathbf{U}_{\boldsymbol{\Psi}}\herm\boldsymbol{\Psi}_{12}\label{eq:Kbar:cf}
\end{equation}
where $\Xi_{\boldsymbol{\Psi}}=2\diag\Bigl(\frac{1}{1+\sqrt{1+4\sigma_{\boldsymbol{\Psi}_{1}}}},\frac{1}{1+\sqrt{1+4\sigma_{\boldsymbol{\Psi}_{2}}}},\ldots,\frac{1}{1+\sqrt{1+4\sigma_{\boldsymbol{\Psi}_{N_{r}}}}}\Bigr)$.
\end{lem}
\begin{IEEEproof}
Please refer to Appendix \ref{sec:proof:closedform:K}.
\end{IEEEproof}
Lemma \ref{lem:optsol:findKbar} implies that $\mathbf{K}_{n}\succ\mathbf{0}$
for all $n$ and thus the $\mathbf{X}$-update is well defined.

To conclude this subsection we note that a similar solution was proposed
in our previous work of \cite{ThangNguyen2020}. However, the method
in \cite{ThangNguyen2020} is a double-loop iterative algorithm. More
precisely, the outer loop was used to approximate the objective in
\eqref{eq:findX} and the inner loop was used to find a saddle-point
of the resulting approximate minimax subproblems. In contrast, the
PBRA is a single-loop iterative algorithm where the maximization over
$\mathbf{X}$ is done exactly.

\subsection{Convergence Analysis}

The convergence of Algorithm \ref{alg:PBRA} is stated in the following
lemma. 
\begin{lem}
\label{lem:PABRconvergence}Let $\{(\mathbf{X}_{n},\mathbf{K}_{n})\}$
be the iterates generated by Algorithm \ref{alg:PBRA}. Then the following
statements hold
\begin{itemize}
\item $f(\mathbf{X}_{n},\mathbf{K}_{n})\geq0$, $f(\mathbf{X}_{n},\mathbf{K}_{n})\geq f(\mathbf{X}_{n+1},\mathbf{K}_{n+1})$
and thus $\{f(\mathbf{X}_{n},\mathbf{K}_{n})\}$ is convergent.
\item $\{(\mathbf{X}_{n},\mathbf{K}_{n})\}$ contains at least a convergent
subsequence.
\item Every limit points of $\{(\mathbf{X}_{n},\mathbf{K}_{n})\}$ is a
saddle point of \eqref{eq:minimax}.
\end{itemize}
\end{lem}
\begin{IEEEproof}
Please refer to Appendix \ref{sec:proof:PABR:convergence}
\end{IEEEproof}
\textcolor{black}{We again note that Algorithm \ref{alg:PBRA} can
find the secrecy capacity but not necessarily the optimal signaling.
To achieve optimal signaling a further bisection search can be employed
in a similar way to \cite[Algorithm 2]{Loyka2020IPC}. More specifically,
after using Algorithm \ref{alg:PBRA} to solve \eqref{eq:minimax},
the secrecy capacity is known. The idea is to carry out a bisection
search over the total transmit power $P_{0}$ (while other power constraints
are fixed) until the obtained saddle point objective approaches the
secrecy capacity up to a given error tolerance. We refer the interested
readers to \cite{Loyka2020IPC} for further details. It is also interesting
to note that in our extensive numerical experiments, both Algorithms
\ref{alg:ADCA} and \ref{alg:ADCA} give the same objective, meaning
that the solution return by Algorithm \ref{alg:ADCA} is indeed the
optimal signaling.}

\section{Numerical Results\label{sec:Numerical-Results}}

In this section we provide numerical results to evaluate the proposed
algorithms. As mentioned previously the SPC only case has been studied
extensively and thus we concentrate on the secrecy capacity of MIMO
WTC for the case of joint SPC and PAPC. We adopt the Kronecker model
in our numerical investigation \cite{Chuah2002,Kermoal2002}. Specifically,
the channel between Alice and Bob $\mathbf{H}_{b}$ is modeled as
$\mathbf{H}_{b}=\tilde{\mathbf{H}}_{b}\mathbf{R}_{b}^{1/2}$, where
$\tilde{\mathbf{H}}_{b}$ is a matrix of i.i.d. complex Gaussian distribution
with zero mean and unit variance and $\mathbf{R}_{b}$ the corresponding
a transmit correlation matrix. Here we adopt the exponential correlation
model whereby $[\mathbf{R}_{b}]_{i,j}=\bigl(re^{j\phi_{b}}\bigr){}^{|i-j|}$
for a given $r\in[0,1]$ and $\phi_{b}\in[0,2\pi)$ \cite{Loyka2001}.
The channel between Alice and Eve is modeled as $\mathbf{H}_{e}=\gamma\tilde{\mathbf{H}}_{e}\mathbf{R}_{e}^{1/2}$
for a given $\gamma>0$ and $\tilde{\mathbf{H}}_{e}$ and $\mathbf{R}_{e}$
are generated in the same way. The purpose of introducing $\gamma$
is to study the secrecy capacity of the MIMO WTC with respect to the
relative average strength of $\mathbf{H}_{b}$ and $\mathbf{H}_{e}$.
The codes of all algorithms in comparison were written in MATLAB and
executed in a 64-bit Windows PC with 16GB RAM and Intel Core i7, 3.20
GHz. Note that since the noise power is normalized to unity and thus
$P_{0}$ is defined to be the signal to noise ratio (SNR) in this
section. The PAPC is set to $P_{i}=1.2P_{0}/N_{t},\forall i=1,2,\ldots,N_{t}$
which makes the joint SPC and PAPC problem nontrivial.

In all simulations results, the parameter $q$ for Algorithm \ref{alg:ADCA}
is taken as $q=5$. The initial point for both Algorithm \ref{alg:ADCA}
and the AO algorithm is taken as $\mathbf{X}_{0}=\frac{P_{0}}{2}\mathbf{I}$
which satisfies both SPC and PAPC. For Algorithm \ref{alg:PBRA} $\mathbf{K}_{0}$
is set to identity.

\subsection{Convergence Results}

\begin{figure}[H]
\textcolor{black}{\centering}

\textcolor{black}{}\subfloat[Joint SPC and PAPC.]{\textcolor{black}{\includegraphics[bb=62bp 558bp 295bp 741bp,width=0.5\columnwidth]{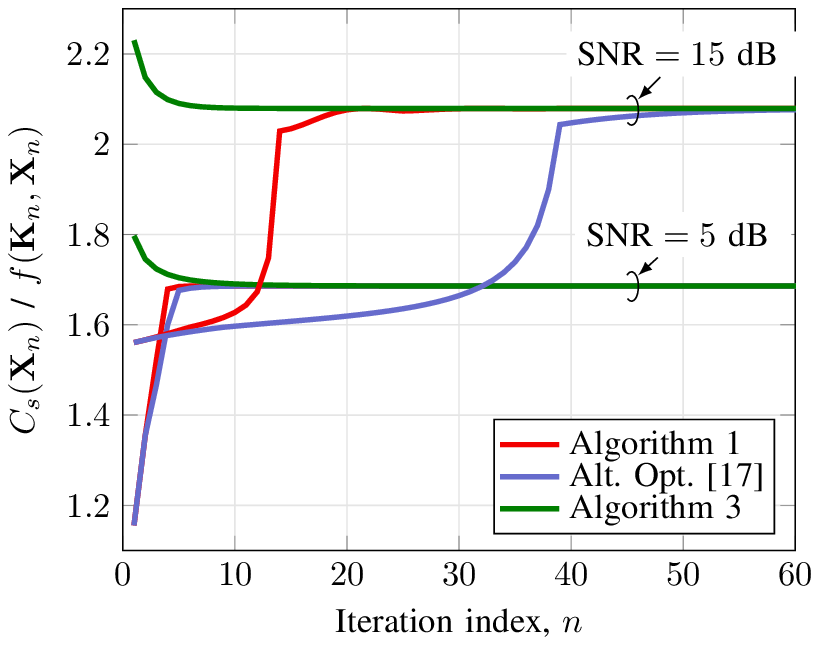}\label{fig:convergence-2}}}\textcolor{black}{}\subfloat[Joint SPC, PAPC and IPC, .]{\textcolor{black}{\includegraphics[bb=62bp 558bp 295bp 741bp,width=0.5\columnwidth]{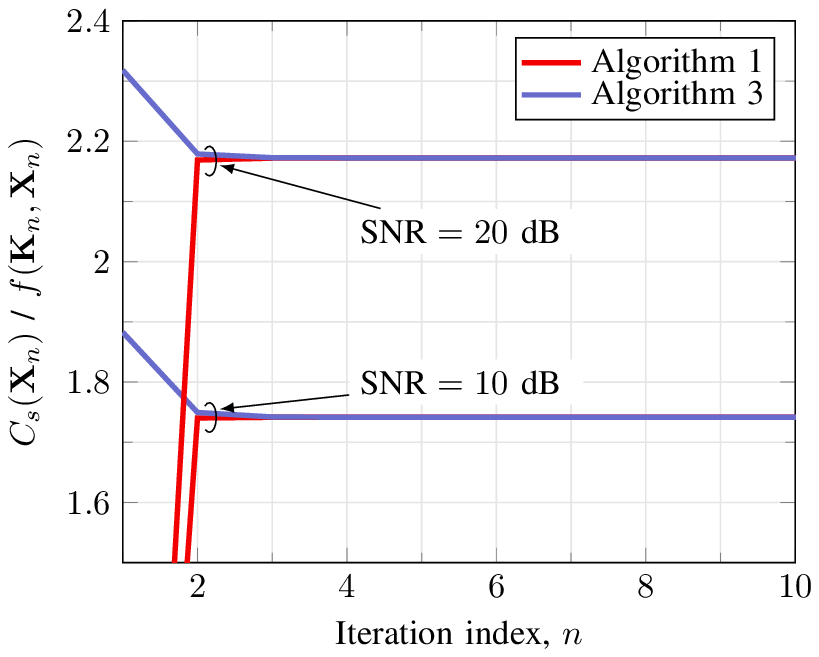}}

\textcolor{black}{\label{fig: allconstraints}}}

\textcolor{black}{\caption{\textcolor{black}{Convergence results of iterative algorithms for
different SNRs and different combination of types of power constraints.}}
\label{fig:correlation-1}}
\end{figure}

\begin{figure}[H]
\centering\includegraphics[bb=62bp 558bp 295bp 741bp,width=0.5\columnwidth]{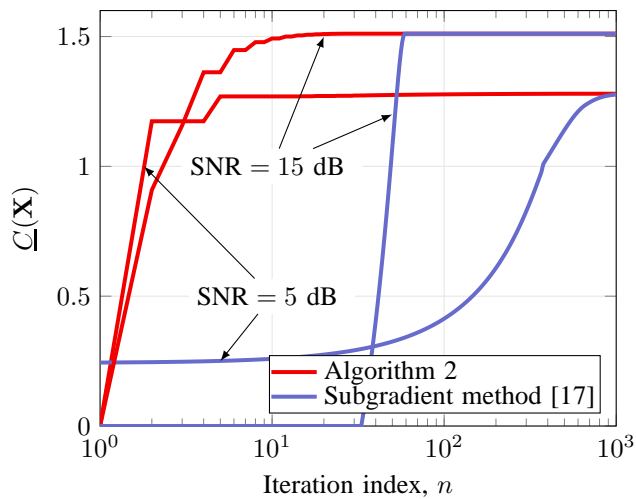}

\caption{Convergence comparison of Algorithm \ref{alg:CoMirror} and the subgradient
method \cite{Li2013} for solving \eqref{eq:DCA:subprob} for different
SNRs.}
\label{fig:convergence:subprob}
\end{figure}

In the first experiment we compare the convergence rate of the proposed
ADCA with the AO method in \cite{Li2013b} for the following channels.

\begin{align*}
\mathbf{H}_{b} & =\begin{bmatrix}-0.3974+j0.5641 &  & -0.0939+j0.2532\\
-0.0216+j0.8051 &  & -0.6734+j0.2605\\
-1.1903-j0.3939 &  & -0.9728-j0.4468\\
0.2017-j0.6897 &  & -0.9450-j0.7306
\end{bmatrix}\\
\mathbf{H}_{e} & =\begin{bmatrix}-0.2015+j0.3127 &  & -0.6178-j1.048\\
-0.0559-j0.3000 &  & -0.3858-j0.2817\\
0.6935+j0.05587 &  & -0.5064-j0.1443
\end{bmatrix}
\end{align*}
which are generated randomly. The convergence rates of the algorithms
in comparison are illustrated in Fig. \ref{fig:convergence-2} for
different values of SNR. For Algorithm \ref{alg:ADCA} and the AO
algorithm we plot the secrecy rate $C_{s}(\mathbf{X}_{n})$ where
$\mathbf{X}_{n}$ is the solution return at the $n$th iteration.
For Algorithm \ref{alg:PBRA} we plot the objective $f(\mathbf{K}_{n},\mathbf{X}_{n})$
in \eqref{eq:minimax}. It can be seen clearly from Fig. \ref{fig:convergence-2}
that Algorithm \ref{alg:ADCA}, being an accelerated version of DCA,
converges much faster than the AO method proposed in \cite{Li2013b},
especially in the high SNR regime. As discussed earlier, since $C_{s}(\mathbf{X})$
is not concave, the extrapolation can be bad which may decrease the
objective. This point is also observed in Fig. \ref{fig:convergence-2}
where Algorithm \ref{alg:ADCA} is not monotonically increasing as
compared to the AO method. Thus, a reasonable stopping criterion for
Algorithm \ref{alg:ADCA} is when the best objective is not improved
during the last, say $10$, iterations. We can also see that $f(\mathbf{K}_{n},\mathbf{X}_{n})$
is indeed an upper bound of $C_{s}(\mathbf{X}_{n})$ and it keeps
decreasing until convergence as expected. It is also interesting,
but not surprising, that at the convergence, all algorithms in comparison
yield a secrecy capacity equal to that return by the minimax barrier
method proposed in \cite{lyoka2015minmaxEQ}, meaning that a globally
optimal solution has been act\textcolor{black}{ually achieved. In
Fig. \ref{fig: allconstraints} we demonstrate}%
\textcolor{black}{{} the convergence rate of Algorithms \ref{alg:ADCA}
and \ref{alg:PBRA} where all three types of constraints:SPC, PAPC
and IPC are included. Particularly, we consider the scenario described
in Example 3 in \cite{Loyka2020IPC}, where Alice also needs to ensure
that the interference to two primary receivers is below a pre-determined
threshold. It is said in \cite{Loyka2020IPC} the involved channels
are non-degraded with negative eigenmodes dominating and are ``hard''
to optimize. More specifically, a Taylor-based algorithm was reported
to be trapped in a bad solution. However this is not the case for
the our proposed ADCA as seen in Fig. \ref{fig: allconstraints}.
The initial point in the ADCA in Fig. \ref{fig: allconstraints} is
the trivial all zero matrix. Surprisingly, our both proposed algorithms
converge very fast to the optimal solution despite the fact that this
case is considered hard to optimize. Although we only show the convergence
of our proposed algorithms to the optimal solution for two representative
values of SNR in Fig. \ref{fig: allconstraints}, our proposed ADCA
indeed always returns the optimal solution for all values of SNRs
considered in \ref{fig: allconstraints}.}
We now demonstrate the usefulness of the convex reformulation of the
secrecy capacity problem for the degraded case. Note that in Fig.
\ref{fig:correlation-1}, the convergence rate of the proposed algorithms
is shown in terms of iteration counts and the per-iteration complexity
is not taken into account. To achieve a more meaningful comparison,
we report in Table \ref{table: Table1Comp} the \emph{average actual
run time} of different methods for solving \eqref{eq:secrecycapacity:org}
when the MIMO WTC is \emph{degraded}, i.e. $\mathbf{H}_{b}\herm\mathbf{H}_{b}\succeq\mathbf{H}_{e}\herm\mathbf{H}_{e}$.
In this case, the power convex solver MOSEK \cite{mosek} can be used
to solve the convex reformulation and is included for comparison.
The average run time in Table \ref{table: Table1Comp} is obtained
from 1000 random channel realizations. For Algorithm \ref{alg:ADCA}
we use Algorithm \ref{alg:CoMirror} to solve the subproblem in each
iteration. Similarly, for Algorithm \ref{alg:PBRA} we modify Algorithm
\ref{alg:CoMirror} to solve \eqref{eq:findX}. The stopping criterion
for Algorithms \ref{alg:ADCA} and \ref{alg:PBRA} i\textcolor{black}{s
when the increase during the last 100 iterations is less than $10^{-5}$.
For comparison purpose, we also report the run time of the barrier
method presented in \cite{lyoka2015minmaxEQ}. It can be seen that
MOSEK is faster to compute the optimal signaling for systems of small
sizes. On the other hand, Algorithm \ref{alg:ADCA} becomes more efficient
when the size of the system increases.}
\begin{table}[tbh]
\textcolor{black}{\caption{\textcolor{black}{Comparison of run-time (in seconds) between the
proposed method for degraded channels \label{table: Table1Comp}}}
}
\centering{}\textcolor{black}{}%
\begin{tabular}{c|>{\centering}p{0.8cm}|c|>{\centering}p{0.8cm}|c}
\hline 
 & \multicolumn{2}{c|}{\textcolor{black}{$\begin{array}{c}
(N_{t},N_{r},N_{e})\\
=(4,4,4)
\end{array}$}} & \multicolumn{2}{c}{\textcolor{black}{$\begin{array}{c}
(N_{t},N_{r},N_{e})\\
=(8,8,8)
\end{array}$}}\tabularnewline
\hline 
\textcolor{black}{\diagbox{Algorithm}{SNR}} & \textcolor{black}{5dB} & \textcolor{black}{10dB} & \textcolor{black}{5dB} & \textcolor{black}{10dB}\tabularnewline
\hline 
\textcolor{black}{MOSEK} & \textbf{\textcolor{black}{0.25}} & \textbf{\textcolor{black}{0.27}} & \textcolor{black}{2.02} & \textcolor{black}{2.01}\tabularnewline
\hline 
\textcolor{black}{Algorithm \ref{alg:ADCA}} & \textcolor{black}{0.9} & \textcolor{black}{1.21} & \textbf{\textcolor{black}{1.43}} & \textbf{\textcolor{black}{1.79}}\tabularnewline
\hline 
\textcolor{black}{Algorithm \ref{alg:PBRA}} & \textcolor{black}{0.88} & \textcolor{black}{1.11} & \textcolor{black}{1.82} & \textcolor{black}{2.81}\tabularnewline
\hline 
\textcolor{black}{Minmax barrier method \cite{lyoka2015minmaxEQ}} & \textcolor{black}{1.47} & \textcolor{black}{1.61} & \textcolor{black}{2.58} & \textcolor{black}{2.68}\tabularnewline
\hline 
\end{tabular}
\end{table}
In the next numerical experiment we compare the convergence rate of
Algorithm \ref{alg:CoMirror} with a subgradient method for solving
\eqref{eq:DCA:subprob}. In particular, we plot the convergence of
both algorithms for the first subproblem (i.e. $n=1$) in Algorithm
\ref{alg:ADCA}. The channels are the same as in Fig. \ref{fig:convergence-2}.
To make relevant reference to Fig. \ref{fig:convergence-2}, we plot
$\underline{C}_{s}(\mathbf{X})$, the lower bound of the secrecy capacity
given by 
\[
\underline{C}_{s}(\mathbf{X})=\bar{f}(\mathbf{X};\mathbf{V}_{0})+\log\bigl|\mathbf{I}+\mathbf{H}_{e}\mathbf{V}_{0}\mathbf{H}_{e}\herm\bigr|-\tr\bigl(\nabla f_{e}(\mathbf{V}_{0})\mathbf{V}_{0}\bigr)\leq C_{s}(\mathbf{X})
\]
instead of the objective in \eqref{eq:DCA:subprob}. That is we include
the constants omitted when deriving \eqref{eq:DCA:subprob}. If $\underline{C}_{s}(\mathbf{X})<0$,
it is replaced by $0$. We note that Algorithm \ref{alg:CoMirror}
and the subgradient method are not monotone in general. To make the
convergence of these two algorithms easier to visualize, we plot the
best $\underline{C}_{s}(\mathbf{X})$ at each iteration. For the subgradient
method we use a constant step size rule. It can be seen clearly that
Algorithm \ref{alg:CoMirror} converges much faster than the subgradient
method in the two considered values of SNR. Furthermore, our extensive
numerical results show that the convergence rate of Algorithm \ref{alg:CoMirror}
is less sensitive to the considered settings and it becomes stabilized
very quickly. We also observe that the convergence of the subgradient
method depends heavily on the choice of the step size. In Fig. \ref{fig:convergence:subprob}
we choose a step size such that the subgradient method produces a
good convergence performance for the considered setting. However,
it cannot guarantee its convergence in many other cases.

\subsection{Impact of Bob-Eve Correlation}

In the next numerical experiment, we investigate the effect of the
channel correlation between Bob and Eve on the achieved MIMO secrecy
capacity. To this end we fix $\phi_{b}=0$ and vary $\phi_{e}$ and
the correlation coefficient $r$ is set to $r=0.9$. We remark that
the transmit correlation matrix of Bob's channel is $\mathbf{R}_{b}=E\{\mathbf{H}_{b}\herm\mathbf{H}_{b}\}$
and of the Eve's channel is $\mathbf{R}_{e}^{\prime}=E\{\mathbf{H}_{e}\herm\mathbf{H}_{e}\}E\{\mathbf{H}_{e}\herm\mathbf{H}_{e}\}=\gamma^{2}\mathbf{R}_{e}$.
The correlation between Bob's channel and Eve's channel can be measured
by the following quantity \cite[Section 3.1.1]{Clerckx2013}
\[
d_{\textrm{corr}}=1-\frac{\tr(\mathbf{R}_{b}\mathbf{R}_{e}^{\prime})}{\bigl\Vert\mathbf{R}_{b}\bigr\Vert_{F}\bigl\Vert\mathbf{R}_{e}^{\prime}\bigr\Vert_{F}}=1-\frac{\tr(\mathbf{R}_{b}\mathbf{R}_{e})}{\bigl\Vert\mathbf{R}_{b}\bigr\Vert_{F}\bigl\Vert\mathbf{R}_{e}\bigr\Vert_{F}}
\]
Note that the $d_{\textrm{corr}}$ is a function of $\phi_{e}$ and
independent of $\gamma$. It is easy to see that if $\mathbf{R}_{b}$
and $\mathbf{R}_{e}$ are identical (apart from a scaling factor),
then $d_{\textrm{corr}}=0$. Roughly speaking, a small value of $d_{\textrm{corr}}$
indicates the two links are highly correlated. On the other hand,
if $d_{\textrm{corr}}$ is close to 1 means that the two links are
highly uncorrelated.

\begin{figure}[H]
\subfloat[$r=0.9$ ]{\includegraphics[bb=62bp 558bp 295bp 741bp,width=0.5\columnwidth]{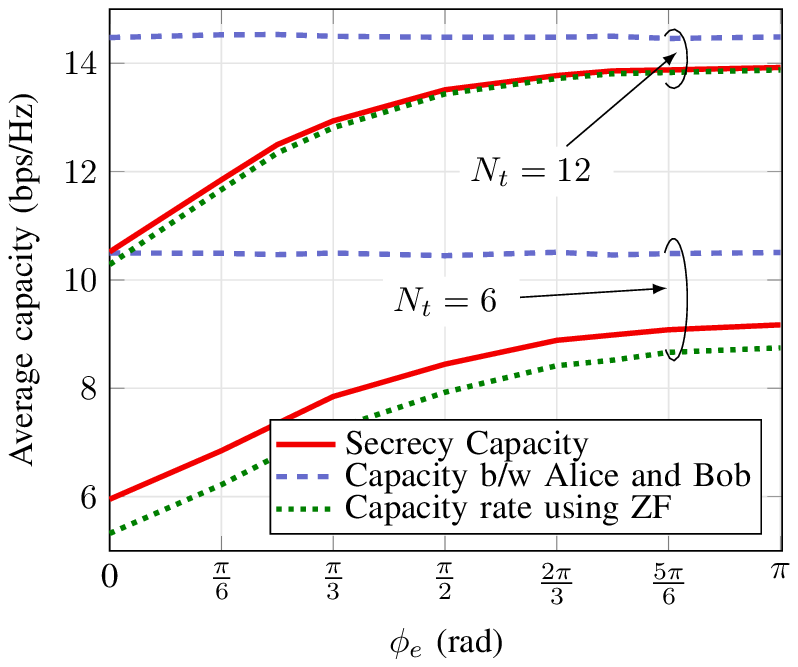}\label{fig:Correlation0.9}}\hfill{}\subfloat[$r=0.2$ ]{\includegraphics[bb=62bp 558bp 295bp 741bp,width=0.5\columnwidth]{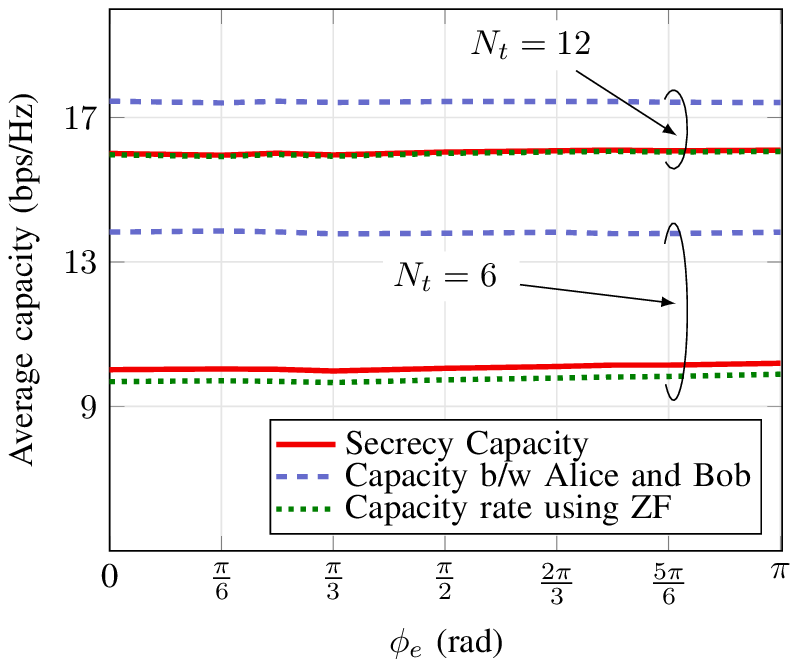}

\label{fig:Correlation0.2}

}

\caption{Effect of channel correlation between Bob and Eve on secrecy capacity
for $N_{r}=4,N_{e}=3$.}
\label{fig:correlation}
\end{figure}

Fig. \ref{fig:correlation} plots the secrecy capacity as a function
of $\phi_{e}$ for two cases of transmit correlation coefficient:
$r=0.9$ (highly correlated antennas) and $r=0.2$ (low correlated
antennas). For comparison purpose, we also include in Fig. \ref{fig:correlation}
the capacity between Alice and Bob when Eve is not present, and the
achieved secrecy rate obtained by the ZF method. Firstly and as expected,
the channel capacity in the absence of Eve is always higher than the
secrecy capacity. However, for highly correlated antennas in Fig.
\ref{fig:Correlation0.9}, the gap is reduced when $\phi_{e}$ is
increased. To explain this, we note that by a direct correlation,
can see that $d_{\textrm{corr}}$ increases from $0$ to 0$.96$ when
$\phi_{e}$ increases from $0$ to $\pi$. Thus, Bob's channel and
Eve's channel become more uncorrelated. Intuitively, we can view $\phi_{e}$
increasing from $0$ to $\pi$ as Eve will move further from Bob along
a circular arc. As a result, Alice can transmit information securely
to Bob through the eigenmodes of $\mathbf{H}_{b}$, without being
comprised by the eigenmodes of $\mathbf{H}_{e}$. That is, the information
leakage is reduced, which in turn increases the secrecy capacity.
Secondly we notice that the secrecy capacity is always higher than
the secrecy capacity rate achieved by the ZF precoding method, and
the gap is also reduced for the same reason as explained above. We
also observe that when the number of antennas at Alice increases from
$N_{t}=6$ to $N_{t}=12$, the gap between the secrecy capacity and
the secrecy achievable rate obtained by the ZF is very marginal. The
reason is that with additional degree of freedom, Alice can now create
multiple beams to Bob without being overheard by Eve.

On the other hand, when the transmit antennas are low correlated as
considered in Fig. \ref{fig:Correlation0.2}, the off-diagonal elements
of both $\mathbf{R}_{b}$ and $\mathbf{R}_{e}$ are very small, compared
to the diagonal elements which are all unity. Thus, both $\mathbf{R}_{b}$
and $\mathbf{R}_{e}$ are very close to the identity matrix and thus
$d_{\textrm{corr}}$ is very small for all considered values of $\phi_{e}$.
Intuitively, Bob's link and Eve's link are highly correlated in this
case. As a result, the gap between the channel capacity in the absence
of Eve and the secrecy capacity is significant and the position of
Eve has little impact on the obtained secrecy capacity. 

\subsection{Impact of Transmit Antenna}

\begin{figure}[tbh]
\subfloat[\textcolor{black}{Power constraints: joint SPC and PAPC}]{\includegraphics[bb=62bp 558bp 295bp 745bp,width=0.5\columnwidth]{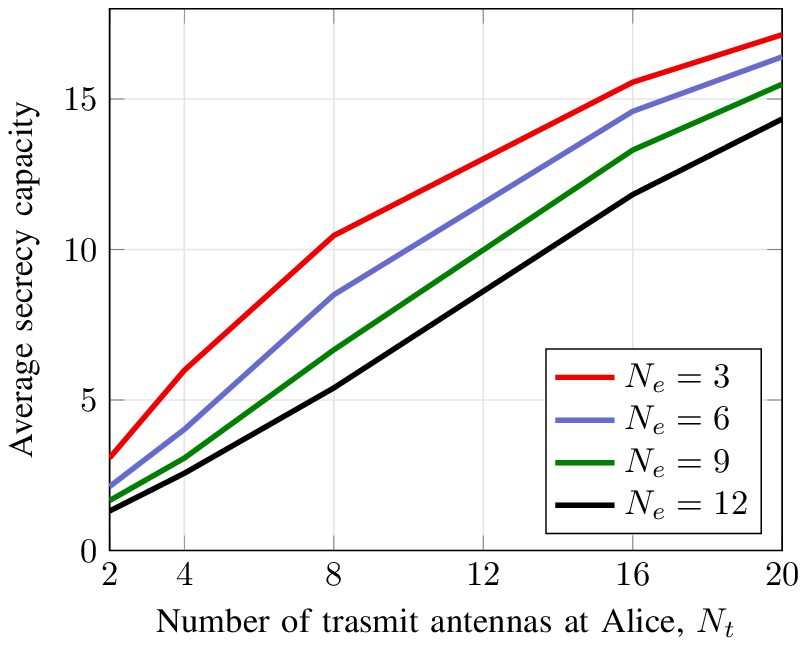}

\label{fig:Tx_Impact-1}}\subfloat[\textcolor{black}{Power constraints: joint SPC, PAPC and IPC. The
number of transmit antennas at the PR is $N_{p}=4$}]{\includegraphics[bb=62bp 558bp 295bp 745bp,width=0.5\columnwidth]{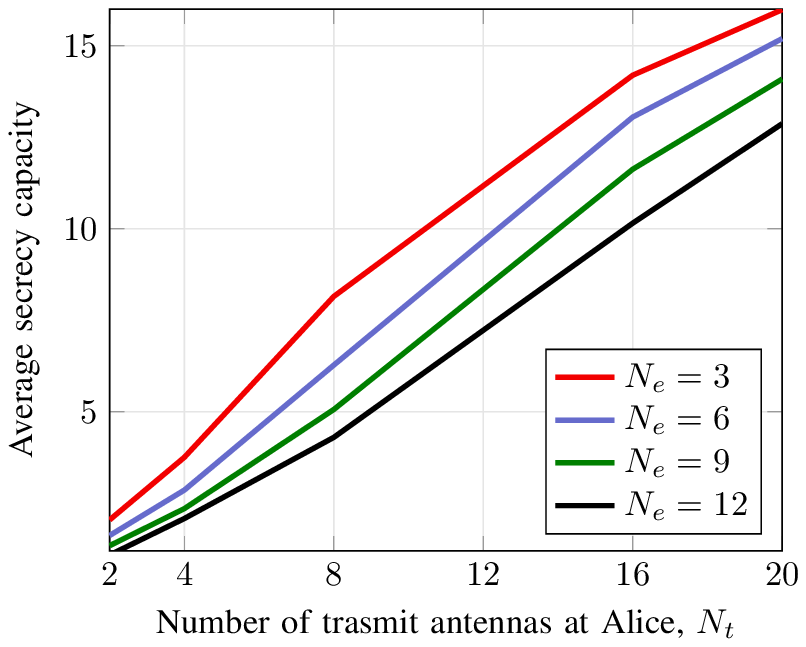}\label{fig:Tx_Impact-2}}

\caption{\textcolor{black}{Secrecy capacity as a function of $N_{t}$ for different
values of $N_{e}$ for different combinations of power constraints.
The number of receiver antennas at Bob is $N_{r}=4$.}}

\label{fig:Tx_Impact}
\end{figure}

We now study how the secrecy capacity scales with the number of transmit
antennas at Alice. Fig. \ref{fig:Tx_Impact} plots the average secrecy
capacity for various numbers of antennas at Eve for different combinations
of power constraints. The number of receiver antennas at Bob is $N_{r}=4$.
The correlation coefficient is set to $r=0.9$ and the parameter $\gamma$
is taken $\gamma=0.9$. For Fig. \ref{fig:Tx_Impact-2} we consider
the scenario where Alice has to limit the interference to a primary
receiver (PR) below a threshold. The channel between Alice and PR
is modeled as $\mathbf{H}_{p}=\gamma\tilde{\mathbf{H}}_{p}\mathbf{R}_{p}^{1/2}$,
where $\tilde{\mathbf{H}}_{p}$ and $\mathbf{R}_{p}$ are generated
in the same way as explained above for Bob's and Eve' channels, i.e.
$[\mathbf{R}_{p}]_{i,j}=\bigl(re^{j\phi_{p}}\bigr){}^{|i-j|}$ where
$\phi_{p}=\frac{\pi}{4}$. The number of antennas at the PR is $N_{p}=4$
and the interference threshold at the PR is $5$ dB. As can be seen
in Fig. \ref{fig:Tx_Impact}, the secrecy capacity increases with
the number of the transmit antennas at Alice, which is expected. Simultaneously,
we also observe that the secrecy capacity is reduced when the number
of antennas at Eve increases. In particular, Eve can significantly
decrease the secrecy capacity when $N_{e}$ is much larger than $N_{t}$.
This is because the null space of $\mathbf{H}_{b}$ will increasingly
intersect with the space spanned by $\mathbf{H}_{e}$. It is also
clear from Fig. \ref{fig:Tx_Impact-2} that including the IPC can
reduce the secrecy rate.

\section{Conclusion}

We have proposed efficient numerical solutions for finding the secrecy
capacity of MIMO WTC subject to joint sum power constraint and per
antenna power constraint. This problem is non-convex in general, and
thus, is difficult to find an optimal solution. Our first contribution
has been a convex reformulation of the secrecy problem for the degraded
MIMO WTC. For non-degraded cases, we have proposed ADCA that solves
the secrecy problem directly and PBRA that solves the equivalent convex-concave
program. In particular, we have also presented the CoMirror algorithm
which efficiently solves the subconvex problems resulting from the
ADCA and the PBRA. We have carried out \textcolor{black}{numerical
experiments to demonstrate the effectiveness of the proposed solutions.
In particular, the convergence rate of the proposed algorithm is much
faster than a known solution. We have also shown that the transmit
antenna correlation at Alice and the number of antennas at Eve have
a significant impact on the secrecy capacity. We note that artificial
noise is a good technique to enhance physical layer security. In this
regard, it is interesting to see if the proposed methods in this paper
can be extended to deal with this technology, which deserves a separate
through study and thus is left for future work.}

\appendix{}

\subsection{\label{sec:lem:convexreform:proof}Proof of Lemma \ref{lem:convexreform}}

To proceed, we first rewrite $C_{s}(\mathbf{X})$ as
\begin{align}
C_{s}(\mathbf{X}) & =\ln\bigl|\mathbf{I}+\mathbf{X}\mathbf{H}_{b}\herm\mathbf{H}_{b}\bigr|-\ln\bigl|\mathbf{I}+\mathbf{X}\mathbf{H}_{e}\herm\mathbf{H}_{e}\bigr|\nonumber \\
 & =\ln\bigl|\mathbf{I}+\mathbf{X}\mathbf{H}_{e}\herm\mathbf{H}_{e}+\mathbf{X}\boldsymbol{\Delta}\bigr|-\ln\bigl|\mathbf{I}+\mathbf{X}\mathbf{H}_{e}\herm\mathbf{H}_{e}\bigr|\nonumber \\
 & =\ln\bigl|\mathbf{I}+\bigl(\mathbf{I}+\mathbf{X}\mathbf{H}_{e}\herm\mathbf{H}_{e}\bigr)^{-1}\mathbf{X}\boldsymbol{\Delta}\bigr|\nonumber \\
 & =\ln\bigl|\mathbf{I}+\boldsymbol{\Delta}^{1/2}\bigl(\mathbf{I}+\mathbf{X}\mathbf{H}_{e}\herm\mathbf{H}_{e}\bigr)^{-1}\mathbf{X}\boldsymbol{\Delta}^{1/2}\bigr|\label{eq:secrate:rewrite}
\end{align}
 where we have used the fact that $\ln|\mathbf{I}+\mathbf{A}\mathbf{B}|=\ln|\mathbf{I}+\mathbf{B}\mathbf{A}|$.
Using the matrix inversion lemma \cite[Fact 2.16.21]{Bernstein:2009},
we have 
\begin{equation}
\bigl(\mathbf{I}+\mathbf{X}\mathbf{H}_{e}\herm\mathbf{H}_{e}\bigr)^{-1}=\mathbf{I}-\mathbf{X}\mathbf{H}_{e}\herm\bigl(\mathbf{I}+\mathbf{H}_{e}\mathbf{X}\mathbf{H}_{e}\herm\bigr)^{-1}\mathbf{H}_{e}
\end{equation}
and thus $C_{s}(\mathbf{X})$ is further equivalently expressed as
\begin{align*}
C_{s}(\mathbf{X}) & =\ln\Bigl|\mathbf{I}+\boldsymbol{\Delta}^{1/2}\bigl(\mathbf{X}-\mathbf{X}\mathbf{H}_{e}\herm\bigl(\mathbf{I}+\mathbf{H}_{e}\mathbf{X}\mathbf{H}_{e}\herm\bigr)^{-1}\mathbf{H}_{e}\mathbf{X}\bigr)\boldsymbol{\Delta}^{1/2}\Bigr|\\
 & =\ln\bigl|\mathbf{I}+\boldsymbol{\Delta}^{1/2}\mathbf{X}\boldsymbol{\Delta}^{1/2}-\boldsymbol{\Delta}^{1/2}\mathbf{X}\mathbf{H}_{e}\herm\bigl(\mathbf{I}+\mathbf{H}_{e}\mathbf{X}\mathbf{H}_{e}\herm\bigr)^{-1}\mathbf{H}_{e}\mathbf{X}\boldsymbol{\Delta}^{1/2}\bigr|.
\end{align*}
The proof is due to some collective results in \cite[Section 3.2]{Bental:2001}.
Let $F(\mathbf{X})=\mathbf{I}+\boldsymbol{\Delta}^{1/2}\mathbf{X}\boldsymbol{\Delta}^{1/2}-\boldsymbol{\Delta}^{1/2}\mathbf{X}\mathbf{H}_{e}\herm\bigl(\mathbf{I}+\mathbf{H}_{e}\mathbf{X}\mathbf{H}_{e}\herm\bigr)^{-1}\mathbf{H}_{e}\mathbf{X}\boldsymbol{\Delta}^{1/2}$
which is a matrix-valued function of $\mathbf{X}$. It is easy to
see that \eqref{eq:secrecycapacity:org} is equivalent to \begin{subequations}
\begin{align}
\underset{\mathbf{X}\succeq\mathbf{0},\mathbf{Y}\succeq\mathbf{0}}{\mathrm{maximize}} & \ \ln|\mathbf{Y}|\\
\st & \ F(\mathbf{X})\succeq\mathbf{Y}\\
 & \ \mathbf{X}\in\mathcal{X}
\end{align}
\end{subequations} which is in fact a ``$\succeq$-epigraph'' form
of \eqref{eq:secrecycapacity:org}. Further note that the constraint
$F(\mathbf{X})\succeq\mathbf{Y}$ is equivalent to 
\[
\mathbf{I}+\boldsymbol{\Delta}^{1/2}\mathbf{X}-\mathbf{Y}-\boldsymbol{\Delta}^{1/2}\mathbf{X}\mathbf{H}_{e}\herm\bigl(\mathbf{I}+\mathbf{H}_{e}\mathbf{X}\mathbf{H}_{e}\herm\bigr)^{-1}\mathbf{H}_{e}\mathbf{X}\boldsymbol{\Delta}^{1/2}\succeq\mathbf{0}
\]
which can be rewritten as \eqref{eq:Slemma} using \cite[Lemma 4.2.1]{Bental:2001}
and thus completes the proof.

\subsection{Proof of Lemma \ref{lem:ADCA:convergence}\label{subsec:proof:ADCA:convergence}}

We adapt the arguments in \cite{NhatPhan2017} for the maximization
context. First note that $\mathbf{X}_{n}$ solves \eqref{eq:DCA:subprob}
and thus we have 
\begin{equation}
f_{b}(\mathbf{X}_{n})-\tr\bigl(\nabla f_{e}(\mathbf{V}_{n-1})\bigr)\mathbf{X}_{n}\geq f_{b}(\mathbf{V}_{n-1})-\tr\bigl(\nabla f_{e}(\mathbf{V}_{n-1})\mathbf{V}_{n-1}\bigr)
\end{equation}
which equivalent to 
\begin{equation}
f_{b}(\mathbf{X}_{n})\geq f_{b}(\mathbf{V}_{n-1})-\tr\bigl(\nabla f_{e}(\mathbf{V}_{n-1})\bigl(\mathbf{V}_{n-1}-\mathbf{X}_{n}\bigr)\bigr)\label{eq:ADCA:proof1}
\end{equation}
The concavity of $f_{e}(\mathbf{X})$ implies
\begin{equation}
f_{e}(\mathbf{X}_{n})\leq f_{e}(\mathbf{V}_{n-1})+\tr\left(\nabla f_{b}(\mathbf{V}_{n-1})\bigl(\mathbf{X}_{n}-\mathbf{V}_{n-1}\bigr)\right)\label{eq:ADCA:proof2}
\end{equation}
Combining \eqref{eq:ADCA:proof1} and \eqref{eq:ADCA:proof2} yields
\begin{equation}
C_{s}(\mathbf{X}_{n})=f_{b}(\mathbf{X}_{n})-f_{e}(\mathbf{X}_{n})\geq f_{b}(\mathbf{V}_{n-1})-f_{e}(\mathbf{V}_{n-1})=C_{s}(\mathbf{V}_{n-1}).
\end{equation}
It follows from Step \ref{alg:ADCA:updateV} of Algorithm \ref{alg:ADCA}
that 
\begin{equation}
C(\mathbf{V}_{n-1})\geq\min\bigl(C_{s}(\mathbf{X}_{n-1}),C_{s}(\mathbf{X}_{n-2}),\ldots,C(\mathbf{X}_{[n-1-q]_{+}})\bigr)=\gamma_{n-1}.
\end{equation}
and thus we obtain
\begin{equation}
C_{s}(\mathbf{X}_{n})\geq C_{s}(\mathbf{V}_{n-1})\geq\gamma_{n-1}.\label{eq:ADCA:proof3}
\end{equation}
Consequently the following inequality also holds \begin{subequations}
\begin{align}
C_{s}(\mathbf{X}_{n+1}) & \geq\gamma_{n}=\min\bigl(C_{s}(\mathbf{X}_{n}),C_{s}(\mathbf{X}_{n-1}),\ldots,C(\mathbf{X}_{[n-q]_{+}})\bigr)\\
 & \geq\min\bigl(C_{s}(\mathbf{X}_{n}),C_{s}(\mathbf{X}_{n-1}),\ldots,C(\mathbf{X}_{n+1-q}),C(\mathbf{X}_{[n-q]_{+}}),C(\mathbf{X}_{[n-1-q]_{+}})\bigr)\\
 & \geq\min\bigl(C_{s}(\mathbf{X}_{n}),\gamma_{n-1}\bigr)=\gamma_{n-1}.
\end{align}
\end{subequations} By repeating this process we can easily see that
\begin{equation}
C_{s}(\mathbf{X}_{n+m})\geq\gamma_{n-1},m=0,1,\ldots,q.\label{eq:ADCA:proof4}
\end{equation}
Therefore we obtain \begin{subequations}
\begin{align}
\gamma_{n+q} & =\min\bigl(C_{s}(\mathbf{X}_{n+q}),C_{s}(\mathbf{X}_{n+q-1}),\ldots,C(\mathbf{X}_{n})\bigr)\\
 & \geq\min\bigl(C_{s}(\mathbf{X}_{n+q}),\gamma_{n-1}\bigr)=\gamma_{n-1}\label{eq:ADCA:proof5}
\end{align}
\end{subequations} where \eqref{eq:ADCA:proof5} follows from \eqref{eq:ADCA:proof4}.

It is easy to check that the sequence $\{\gamma_{n}\}$ is bounded
above and thus is convergent. Also, since the feasible set $\mathcal{X}$
is compact and convex, there exist a convergent subsequence. Let $\text{\ensuremath{\mathbf{X}_{n_{j}}}}$
be the subsequence converging to a limit point $\mathbf{X}^{\ast}$.
Without loss of generality, we assume that $\mathbf{V}_{n_{j}-1}$
converges to a limit point $\mathbf{V}^{\ast}$. Due to the strict
concavity and continuity of the objective in each subproblem, it must
follow that $\mathbf{X}^{\ast}=\mathbf{V}^{\ast}$. Since $\mathbf{X}_{n_{j}}$
is the solution to \eqref{eq:DCA:subprob} we have 
\begin{equation}
\tr\bigl(\nabla f_{b}(\mathbf{X}_{n_{j}})-\nabla f_{e}(\mathbf{V}_{n_{j}-1})\bigr)\bigl(\mathbf{X}-\mathbf{X}_{n_{j}}\bigr)\leq0,\forall\mathbf{X}\in\mathcal{X}.
\end{equation}
Let $j\to\infty$ give $\tr\bigl(\nabla f_{b}(\mathbf{X}^{\ast})-\nabla f_{e}(\mathbf{X}^{\ast})\bigr)\bigl(\mathbf{X}-\mathbf{X}^{\ast}\bigr)\leq0,\forall\mathbf{X}\in\mathcal{X}$,
meaning that $\mathbf{X}^{\ast}$ is a critical point of \eqref{eq:secrecycapacity:org}
and thus completes the proof.

\subsection{\label{sec:convergence:comirror}Convergence Analysis of Algorithm
\ref{alg:CoMirror}}

Algorithm \ref{alg:CoMirror} is a special case of the CoMirror Algorithm
in \cite{Beck2010} when the distance generating function (also known
as the kernel function of the mirror map) \cite{Juditsky2011} is
chosen as $\Theta(\mathbf{X})=\frac{1}{2}\left\Vert \mathbf{X}\right\Vert _{F}^{2}$
and the associated norm over $\mathcal{X}_{\mathrm{spc}}$ is the
Frobenius norm. For a given initial point $\mathbf{X}^{0}\in\mathcal{X}_{\textrm{spc}}$,
the parameter $\Omega$ is defined as
\[
\Omega=\frac{1}{\sqrt{2}}\underset{\mathbf{X}\in\mathcal{X}_{\textrm{spc}}}{\max}\ \bigl\Vert\mathbf{X}-\mathbf{X}^{0}\bigr\Vert_{F}.
\]
If $\mathbf{X}^{0}=\mathbf{0}$, then it is easy to see that $\Omega\leq\frac{1}{\sqrt{2}}P_{0}$.
In general we can find of an upper bound of $\Omega$ as 
\begin{align*}
\Omega & =\frac{1}{\sqrt{2}}\underset{\mathbf{X}\in\mathcal{X}_{\textrm{spc}}}{\max}\ \bigl\Vert\mathbf{X}-\mathbf{X}^{0}\bigr\Vert_{F}=\frac{1}{\sqrt{2}}\underset{\mathbf{X}\in\mathcal{X}_{\textrm{spc}}}{\max}\ \bigl\Vert\mathbf{X}-\mathbf{X}^{0}\bigr\Vert_{F}\\
 & =\frac{1}{\sqrt{2}}\underset{\mathbf{X}\in\mathcal{X}_{\textrm{spc}}}{\max}\sqrt{\bigl\Vert\mathbf{X}\bigr\Vert_{F}^{2}-2\tr\bigl(\mathbf{X}^{0}\mathbf{X}\bigr)+\bigl\Vert\mathbf{X}^{0}\bigr\Vert_{F}^{2}}\\
 & \leq\sqrt{\frac{P_{0}^{2}+\bigl\Vert\mathbf{X}^{0}\bigr\Vert_{F}^{2}}{2}}.
\end{align*}
The next step is show that $\nabla g_{m}(\mathbf{X})=\diag(\mathbf{e}_{m})$,
where $m=\underset{1\leq i\leq N_{t}}{\arg\max}g_{i}(\mathbf{X})$
is a subgradient of the pointwise maximum $g(\mathbf{X})=\underset{1\leq i\leq N_{t}}{\max}g_{i}(\mathbf{X})$.
This is a trivial result since $g_{m}(\mathbf{X})$ is an active function
at $\mathbf{X}$ (i.e. $g(\mathbf{X})=g_{m}(\mathbf{X})$). It is
also easy to see that $\bigl\Vert\nabla g_{m}(\mathbf{X})\bigr\Vert_{F}\leq1$
and thus is bounded. The final step to establish the convergence of
Algorithm \ref{alg:CoMirror} is to show that $\bigl\Vert\nabla\bar{f}(\mathbf{X})\bigr\Vert_{F}$
is bounded. To this end, we recall that 
\begin{equation}
\nabla_{\mathbf{X}}\bar{f}(\mathbf{X})=\mathbf{H}_{b}\herm(\mathbf{I}+\mathbf{H}_{b}\mathbf{X}\mathbf{H}_{b}\herm)^{-1}\mathbf{H}_{b}-\boldsymbol{\Gamma}_{n-1}\label{eq:gradfX}
\end{equation}
and thus
\begin{align}
\left\Vert \nabla_{\mathbf{X}}\bar{f}(\mathbf{X})\right\Vert _{F} & \leq\left\Vert \boldsymbol{\Gamma}_{n-1}\right\Vert _{F}+\bigl\Vert\mathbf{H}_{b}\herm(\mathbf{I}+\mathbf{H}_{b}\mathbf{X}\mathbf{H}_{b}\herm)^{-1}\mathbf{H}_{b}\bigr\Vert_{F}\nonumber \\
 & \leq\left\Vert \boldsymbol{\Gamma}_{n-1}\right\Vert _{F}+\bigl\Vert\mathbf{H}_{b}\herm\mathbf{H}_{b}\bigr\Vert_{F}\label{eq:beta12_2}
\end{align}
which holds because $(\mathbf{I}+\mathbf{H}_{b}\mathbf{X}\mathbf{H}_{b}\herm)^{-1}\preceq\mathbf{I}$
for $\mathbf{X}\succeq\mathbf{0}$.

\subsection{Proof of Lemma \ref{lem:optsol:findKbar}\label{sec:proof:closedform:K}}

A very brief proof of Lemma \ref{lem:optsol:findKbar} was provided
in \cite{ThangNguyen2020}. Herein we present a more rigorous proof.
The idea is based on manipulating the Karush-Kuhn-Tucker (KKT) conditions
of problem \ref{eq:findKbar}. Since problem \eqref{eq:minKbar} is
convex and strong duality holds, KKT conditions are necessary and
sufficient for an optimal solution. Let $\mathbf{Z}$ be the Lagrangian
multiplier for the constraint $\mathbf{I}-\bar{\mathbf{K}}\bar{\mathbf{K}}\herm\succeq\mathbf{0}$.
Then the KKT conditions of \eqref{eq:minKbar} are given by\begin{subequations}
\begin{align}
\boldsymbol{\Psi}_{12}+(\mathbf{I}-\bar{\mathbf{K}}\bar{\mathbf{K}}\herm)^{-1}\bar{\mathbf{K}}+\mathbf{Z}\bar{\mathbf{K}} & =0\label{eq:KbarKKT1}\\
\mathbf{I}-\bar{\mathbf{K}}\bar{\mathbf{K}}\herm & \succeq\mathbf{0}\label{eq:KbarKKT2}\\
\mathbf{Z} & \succeq\mathbf{0}\label{eq:KbarKKT3}\\
\tr\left(\left(\mathbf{I}-\bar{\mathbf{K}}\bar{\mathbf{K}}\herm\right)\mathbf{Z}\right) & =\mathbf{0}\label{eq:Kbar:KKT4}
\end{align}
\end{subequations} where we have used the results in \cite{Hjorungens2007matrixAlgebra}
to obtain \eqref{eq:KbarKKT1}. Let us assume for the moment that
$\mathbf{I}-\bar{\mathbf{K}}\bar{\mathbf{K}}\herm\succ\mathbf{0}$.
Then it follows immediately from \eqref{eq:Kbar:KKT4} that $\mathbf{Z}=\mathbf{0}$
and thus we have 
\begin{equation}
(\mathbf{I}-\bar{\mathbf{K}}\bar{\mathbf{K}}\herm)^{-1}\bar{\mathbf{K}}=-\boldsymbol{\Psi}_{12}\label{eq:KKT:Kbar}
\end{equation}
which yields
\begin{equation}
(\mathbf{I}-\bar{\mathbf{K}}\bar{\mathbf{K}}\herm)^{-1}\bar{\mathbf{K}}\bar{\mathbf{K}}\herm(\mathbf{I}-\bar{\mathbf{K}}\bar{\mathbf{K}}\herm)^{-1}=\boldsymbol{\Psi}_{12}\boldsymbol{\Psi}_{12}\herm.\label{eq:KKT:EVD}
\end{equation}
Let $\bar{\mathbf{K}}\bar{\mathbf{K}}\herm=\mathbf{U}_{\bar{\mathbf{K}}}\bar{\boldsymbol{\Sigma}}_{\bar{\mathbf{K}}}\mathbf{U}_{\bar{\mathbf{K}}}^{\dagger}$
and $\boldsymbol{\Psi}_{12}\boldsymbol{\Psi}_{12}\herm=\mathbf{U}_{\mathbf{\boldsymbol{\Psi}}}\bar{\boldsymbol{\Sigma}}_{\mathbf{\mathbf{\boldsymbol{\Psi}}}}\mathbf{U}_{\mathbf{\boldsymbol{\Psi}}}^{\dagger}$
be the eigenvalue decomposition of $\mathbf{\bar{\mathbf{K}}\bar{\mathbf{K}}\herm}$
and $\boldsymbol{\Psi}\boldsymbol{\Psi}\herm$, respectively, where
$\mathbf{U}_{\bar{\mathbf{K}}}\in\mathbb{C}^{N_{r}\times N_{r}}$
and $\mathbf{U}_{\boldsymbol{\Psi}}\in\mathbb{C}^{N_{r}\times N_{r}}$
are unitary and $\bar{\boldsymbol{\Sigma}}_{\bar{\mathbf{K}}}=\diag(\sigma_{\bar{\mathbf{K}}_{1}},\sigma_{\bar{\mathbf{K}}_{2}},\ldots,\sigma_{\bar{\mathbf{K}}_{N_{r}}})$
$\bar{\boldsymbol{\Sigma}}_{\mathbf{\mathbf{\boldsymbol{\Psi}}}}=\diag(\sigma_{\boldsymbol{\Psi}_{1}},\sigma_{\boldsymbol{\Psi}_{2}},\ldots,\sigma_{\boldsymbol{\Psi}_{N_{r}}})$.
Note that $\sigma_{\bar{\mathbf{K}}_{i}}$and $\sigma_{\boldsymbol{\Psi}_{i}}$
are the eigenvalues of $\bar{\mathbf{K}}\bar{\mathbf{K}}\herm$ and
$\boldsymbol{\Psi}\boldsymbol{\Psi}\herm$, respectively. Then \eqref{eq:KKT:EVD}
is equivalent to
\begin{equation}
\mathbf{U}_{\bar{\mathbf{K}}}(\mathbf{I}-\bar{\boldsymbol{\Sigma}}_{\bar{\mathbf{K}}})^{-1}\bar{\boldsymbol{\Sigma}}_{\bar{\mathbf{K}}}(\mathbf{I}-\bar{\boldsymbol{\Sigma}}_{\bar{\mathbf{K}}})^{-1}\mathbf{U}_{\bar{\mathbf{K}}}^{\dagger}=\mathbf{U}_{\boldsymbol{\Psi}}\bar{\boldsymbol{\Sigma}}_{\boldsymbol{\Psi}}\mathbf{U}_{\boldsymbol{\Psi}}^{\dagger}.\label{eq:modifiedSVDKbar}
\end{equation}
Thus we can set\begin{subequations}
\begin{align}
\mathbf{U}_{\bar{\mathbf{K}}} & =\mathbf{U}_{\boldsymbol{\Psi}}\\
(\mathbf{I}-\bar{\boldsymbol{\Sigma}}_{\bar{\mathbf{K}}})^{-1}\bar{\boldsymbol{\Sigma}}_{\bar{\mathbf{K}}}(\mathbf{I}-\bar{\boldsymbol{\Sigma}}_{\bar{\mathbf{K}}})^{-1} & =\bar{\boldsymbol{\Sigma}}_{\boldsymbol{\Psi}}\label{eq:KKT:sigma}
\end{align}
\end{subequations} and the objective is to find $\bar{\boldsymbol{\Sigma}}_{\bar{\mathbf{K}}}$
such that \eqref{eq:KKT:sigma} is satisfied. It is easy to see that
\eqref{eq:modifiedSVDKbar} gives
\begin{equation}
\frac{\sigma_{\bar{\mathbf{K}}_{i}}}{(1-\sigma_{\bar{\mathbf{K}}_{i}})^{2}}=\sigma_{\boldsymbol{\Psi}_{i}},i=1,2,\ldots,N_{r}\label{eq:findK_quard}
\end{equation}
 Solving for $\sigma_{\bar{\mathbf{K}}_{i}}$yields
\begin{equation}
\sigma_{\bar{\mathbf{K}}_{i}}=\begin{cases}
0 & \sigma_{\Psi_{i}}=0\\
0.5\left\{ \left(2+\frac{1}{\sigma_{\boldsymbol{\Psi}_{i}}}\right)-\sqrt{\left(2+\frac{1}{\sigma_{\boldsymbol{\Psi}_{i}}}\right)^{2}-4}\right\}  & \sigma_{\Psi_{i}}>0.
\end{cases}
\end{equation}
We remark that $1>\sigma_{\bar{\mathbf{K}}_{i}}$, $\forall i=1,2,\ldots,N_{r}$
and thus $\mathbf{I}-\bar{\mathbf{K}}\bar{\mathbf{K}}\herm\succ\mathbf{0}$
as assumed above and it satisfies the KKT conditions. After some algebraic
steps we simplify the above equation as
\begin{equation}
\sigma_{\bar{\mathbf{K}}_{i}}=\frac{4\sigma_{\boldsymbol{\Psi}_{i}}}{\bigl(1+\sqrt{1+4\sigma_{\boldsymbol{\Psi}_{i}}}\bigr)^{2}}
\end{equation}
 and thus
\[
\mathbf{I}-\bar{\mathbf{K}}\bar{\mathbf{K}}\herm=2\mathbf{U}_{\bar{\mathbf{K}}}\diag\biggl(\frac{1}{1+\sqrt{1+4\sigma_{\boldsymbol{\Psi}_{1}}}},\frac{1}{1+\sqrt{1+4\sigma_{\boldsymbol{\Psi}_{2}}}},\ldots,\frac{1}{1+\sqrt{1+4\sigma_{\boldsymbol{\Psi}_{N_{r}}}}}\biggr)\mathbf{U}_{\bar{\mathbf{K}}}\herm.
\]
Multiplying both sides of \eqref{eq:KKT:Kbar} with $\mathbf{I}-\bar{\mathbf{K}}\bar{\mathbf{K}}\herm$
and using the above equation results in which completes the proof.

\subsection{Proof of Lemma \ref{lem:PABRconvergence} \label{sec:proof:PABR:convergence}}

First we note that for a given $\mathbf{K}_{n}$, $\mathbf{X}_{n}$
is the capacity achieving covariance matrix of the combined MIMO channel
that contains both $\mathbf{H}_{b}$ and $\mathbf{H}_{e}$ where $\mathbf{K}_{n}$
is the effective noise \cite{Oggier2011SecCapEq}. Thus, $f(\mathbf{X}_{n},\mathbf{K}_{n})$
is always non-negative $f(\mathbf{X}_{n},\mathbf{K}_{n})\geq0$. The
main idea behind the proof of the monotonic decrease of the objective
sequence $\{f(\mathbf{K}_{n},\mathbf{X}_{n})\}$ is to exploit the
fact that the term $\log|\mathbf{K}+\mathbf{H}\mathbf{X}\mathbf{H}\herm|-\log|\mathbf{I}+\mathbf{H}_{e}\mathbf{X}\mathbf{H}_{e}\herm|$
is \emph{jointly concave} with $\mathbf{K}$ and $\mathbf{X}$. In
this regard, the following inequality is straightforward
\begin{align}
f(\mathbf{K},\mathbf{X}) & \leq\log|\mathbf{K}_{n}+\mathbf{H}\mathbf{X}_{n}\mathbf{H}\herm|+\tr(\boldsymbol{\Psi}_{n}(\mathbf{K}-\mathbf{K}_{n}))\nonumber \\
 & \quad+\tr(\boldsymbol{\Phi}_{n}(\mathbf{X}-\mathbf{X}_{n}))-\log|\mathbf{K}|-\log|\mathbf{I}+\mathbf{H}_{e}\mathbf{X}_{n}\mathbf{H}_{e}\herm|,\forall\mathbf{K}\in\mathcal{K}.\label{eq:minmaxobj:UB}
\end{align}
where $\boldsymbol{\Phi}_{n}=\mathbf{H}\herm\bigl(\mathbf{K}_{n}+\mathbf{H}\mathbf{X}_{n}\mathbf{H}\herm\bigr)^{-1}\mathbf{H}-\mathbf{H}_{e}\herm\bigl(\mathbf{I}+\mathbf{H}_{e}\mathbf{X}_{n}\mathbf{H}_{e}\herm\bigr)^{-1}\mathbf{H}_{e}$.
The above inequality is nothing but an affine approximation of $\log|\mathbf{K}+\mathbf{H}\mathbf{X}\mathbf{H}\herm|-\log|\mathbf{I}+\mathbf{H}_{e}\mathbf{X}\mathbf{H}_{e}\herm$
around the point $(\mathbf{X}_{n},\mathbf{K}_{n})$. Substituting
$(\mathbf{K},\mathbf{X}):=(\mathbf{K}_{n+1},\mathbf{X}_{n+1})$ into
\eqref{eq:minmaxobj:UB} we obtain
\begin{align*}
f(\mathbf{K}_{n+1},\mathbf{X}_{n+1}) & \leq\log|\mathbf{K}_{n}+\mathbf{H}\mathbf{X}_{n}\mathbf{H}\herm|-\log\bigl(\mathbf{K}_{n+1}\bigr)\\
 & \hspace{-1cm}-\log|\mathbf{I}+\mathbf{H}_{e}\mathbf{X}_{n}\mathbf{H}_{e}\herm|+\tr(\boldsymbol{\Psi}_{n}(\mathbf{K}_{n+1}-\mathbf{K}_{n}))+\tr(\boldsymbol{\Phi}_{n}(\mathbf{X}_{n+1}-\mathbf{X}_{n}))
\end{align*}
Since $\mathbf{X}_{n}$ is the solution to \eqref{eq:findX} the first
order optimality condition implies
\begin{equation}
\tr\bigl(\boldsymbol{\Phi}_{n}\bigl(\mathbf{X}-\mathbf{X}_{n}\bigr)\bigr)\leq0,\forall\mathbf{X}\in\mathcal{X}.
\end{equation}
Substituting $\mathbf{X}$ by $\mathbf{X}_{n+1}$ yields
\begin{equation}
\tr\Bigl(\boldsymbol{\Phi}_{n}\bigl(\mathbf{X}_{n+1}-\mathbf{X}_{n}\bigr)\leq0,
\end{equation}
and thus
\begin{multline}
f(\mathbf{K}_{n+1},\mathbf{X}_{n+1})\leq\log|\mathbf{K}_{n}+\mathbf{H}\mathbf{X}_{n}\mathbf{H}\herm|-\log|\mathbf{K}_{n+1}|\\
-\log|\mathbf{I}+\mathbf{H}_{e}\mathbf{X}_{n}\mathbf{H}_{e}\herm|+\tr(\boldsymbol{\Psi}_{n}(\mathbf{K}_{n+1}-\mathbf{K}_{n})).\label{eq:descent1}
\end{multline}
 Next we will turn our attention to the $\mathbf{K}$ update. Since
$\mathbf{K}_{n+1}$ solves \eqref{eq:findK}, we have 
\begin{equation}
\tr(\boldsymbol{\Psi}_{n}\mathbf{K}_{n+1})-\log|\mathbf{K}_{n+1}|\leq\tr(\boldsymbol{\Psi}_{n}\mathbf{K})-\log|\mathbf{K}|,\forall\mathbf{K}\in\mathcal{K}
\end{equation}
which is true due to the fact that the optimal objective is less than
or equal to the objective of any feasible solution. Substituting $\mathbf{K}:=\mathbf{K}_{n}$
into the above inequality gives
\begin{equation}
\tr\bigl(\boldsymbol{\Psi}_{n}\mathbf{K}_{n+1}\bigr)-\log\bigl|\mathbf{K}_{n+1}\bigr|\leq\tr(\boldsymbol{\Psi}_{n}\mathbf{K}_{n})-\log\bigl|\mathbf{K}_{n}\bigr|,\forall\mathbf{K}\in\mathcal{K}
\end{equation}
which is equivalent to
\begin{equation}
\tr\bigl(\boldsymbol{\Psi}_{n}\bigl(\mathbf{K}_{n+1}-\mathbf{K}_{n}\bigr)\bigr)-\log\bigl|\mathbf{K}_{n+1}\bigr|\leq-\log\bigl|\mathbf{K}_{n}\bigr|\label{eq:descent2}
\end{equation}
We note that the above inequality is \emph{strict} if $\mathbf{K}_{n+1}\neq\mathbf{K}_{n}$.
Combining \eqref{eq:descent1} and \eqref{eq:descent2} we obtain
\begin{align*}
f(\mathbf{K}_{n+1},\mathbf{X}_{n+1}) & \leq\log|\mathbf{K}_{n}+\mathbf{H}\mathbf{X}_{n}\mathbf{H}\herm|-\log\bigl|\mathbf{K}_{n}\bigr|-\log|\mathbf{I}+\mathbf{H}_{e}\mathbf{X}_{n}\mathbf{H}_{e}\herm|=f(\mathbf{K}_{n},\mathbf{X}_{n}).
\end{align*}
\bibliographystyle{IEEEtran}
\bibliography{IEEEabrv,paper}

\end{document}